\documentclass[aps,prl,reprint,superscriptaddress,longbibliography]{revtex4-2}
\usepackage{braket}
\usepackage[T1]{fontenc}
\usepackage[utf8]{inputenc}
\usepackage{textcomp}

\DeclareUnicodeCharacter{03BD}{\ensuremath{\nu}}
\DeclareUnicodeCharacter{03C9}{\ensuremath{\omega}}
\DeclareUnicodeCharacter{2126}{\ensuremath{\Omega}}
\DeclareUnicodeCharacter{2194}{\ensuremath{\leftrightarrow}}
\DeclareUnicodeCharacter{2212}{\ensuremath{-}}
\DeclareUnicodeCharacter{00B3}{\ensuremath{^3}}
\DeclareUnicodeCharacter{00BD}{\ensuremath{\frac{1}{2}}}
\DeclareUnicodeCharacter{2092}{\ensuremath{_o}}
\DeclareUnicodeCharacter{209B}{\ensuremath{_s}}
\DeclareUnicodeCharacter{209C}{\ensuremath{_t}}
\DeclareUnicodeCharacter{00DF}{\ss}
\DeclareUnicodeCharacter{00E1}{\'a}
\DeclareUnicodeCharacter{00E4}{\"a}
\DeclareUnicodeCharacter{00E8}{\`e}
\DeclareUnicodeCharacter{00E9}{\'e}
\DeclareUnicodeCharacter{00ED}{\'i}
\DeclareUnicodeCharacter{00F3}{\'o}
\DeclareUnicodeCharacter{00F6}{\"o}
\DeclareUnicodeCharacter{00F8}{\o}
\DeclareUnicodeCharacter{00FC}{\"u}
\DeclareUnicodeCharacter{0107}{\'c}
\DeclareUnicodeCharacter{010D}{\v{c}}
\DeclareUnicodeCharacter{011F}{\u{g}}
\usepackage{amsmath,amssymb,bm}
\usepackage{graphicx}
\usepackage{xcolor}
\usepackage{hyperref}
\hypersetup{colorlinks=true,linkcolor=blue,citecolor=blue,urlcolor=blue}

\newcommand{\NiV}{NiV$^{-}$}
\newcommand{\NV}{NV$^{-}$}
\newcommand{\SiV}{SiV$^{-}$}
\newcommand{\SnV}{SnV$^{-}$}
\newcommand{\GeV}{GeV$^{-}$}
\newcommand{\PbV}{PbV$^{-}$}

\begin{document}

\title{A transition-metal qubit in diamond with all-optical control and millisecond quantum memory}

\author{I. M. Morris}
\thanks{These authors contributed equally to this work.}
\affiliation{Michigan State University, Department of Physics and Astronomy; East Lansing, USA}
\affiliation{Michigan State University, Department of Electrical and Computer Engineering; East Lansing, USA}

\author{T. Alberth}
\thanks{These authors contributed equally to this work.}
\affiliation{Michigan State University, Department of Physics and Astronomy; East Lansing, USA}
\affiliation{Michigan State University, Department of Electrical and Computer Engineering; East Lansing, USA}

\author{L. Crooks}
\affiliation{Michigan State University, Department of Physics and Astronomy; East Lansing, USA}
\affiliation{Michigan State University, Department of Electrical and Computer Engineering; East Lansing, USA}

\author{T. L\"uhmann}
\affiliation{University of Leipzig, Felix Bloch Institute for Solid State Physics; Leipzig, Germany}

\author{D. J. Twitchen}
\affiliation{Element Six Global Innovation Centre; Didcot, United Kingdom}

\author{S. Pezzagna}
\affiliation{University of Leipzig, Felix Bloch Institute for Solid State Physics; Leipzig, Germany}

\author{J. Meijer}
\affiliation{University of Leipzig, Felix Bloch Institute for Solid State Physics; Leipzig, Germany}

\author{S. S. Nicley}
\affiliation{Michigan State University, Department of Electrical and Computer Engineering; East Lansing, USA}
\affiliation{Michigan State University, Department of Chemical Engineering and Materials Science; East Lansing, USA}
\affiliation{Coatings and Diamond Technologies Division, Center Midwest (CMW), Fraunhofer USA Inc.; East Lansing, USA}

\author{J. N. Becker}
\email{becke183@msu.edu}
\affiliation{Michigan State University, Department of Physics and Astronomy; East Lansing, USA}
\affiliation{Coatings and Diamond Technologies Division, Center Midwest (CMW), Fraunhofer USA Inc.; East Lansing, USA}

\date{\today}

\begin{abstract}
Quantum networks require qubits that combine efficient optical access, coherent control, and long-lived quantum memory, but realizing all three in one scalable platform remains a central bottleneck. Diamond color centers are leading candidates, yet widely studied defects retain tradeoffs among these capabilities. Here, we show that transition-metal defects in diamond provide a distinct route beyond these platforms by combining spin-orbit-protected ground-state coherence, all-optical control, and near-infrared emission. Using a single nickel-vacancy (\NiV), we demonstrate an all-optically controlled diamond spin qubit with coherence exceeding one millisecond at 1.65 K, compatible with compact closed-cycle cryogenics. We implement Raman Rabi oscillations and Ramsey interferometry and use all-optical dynamical decoupling to extend coherence from $T_2^*$ = 371 ns to $T_2^{\mathrm{CPMG-4}}$ = 1.27 ms, establishing \NiV as a deployable diamond spin-photon interface.
\end{abstract}

\maketitle

A central goal of quantum information science is the realization of quantum networks that distribute entanglement across remote quantum devices \cite{ref1,ref2}. Such networks could enable provably secure communication \cite{ref3,ref4}, modular and scalable quantum computing \cite{ref5,ref6}, and enhanced quantum sensing \cite{ref7,ref8}. Their implementation requires spin-photon interfaces, where a local electron or nuclear spin stores quantum information as it can remain coherent for long times, while photons carry that information between distant nodes rapidly and with low loss or added noise as they interact only weakly with their environment. Practical interfaces must therefore combine efficient optical emission, coherent spin control, and long-lived memory. Diamond color centers are leading candidates because their optical transitions can interface localized electron or nuclear spins with photons \cite{ref9}. The negatively charged nitrogen-vacancy (\NV) center is the prototypical example and has enabled remote spin-spin entanglement \cite{ref10}, loophole-free Bell tests \cite{ref11} and multi-node quantum networks \cite{ref12}, including a recent metropolitan-scale demonstration \cite{ref13}. These advances rely on the \NV center's long spin coherence, even at room temperature, and efficient microwave control of its ground state spin. \cite{ref10,ref14}. However, its low emission into the coherent zero-phonon line (ZPL) \cite{ref15} and lack of inversion symmetry, which leaves its optical transitions susceptible to electric-field-induced spectral diffusion \cite{ref16,ref17}, limit the rate and scalability of optically mediated entanglement generation \cite{ref18}.

\begin{figure*}[t]
    \centering
    \includegraphics[width=0.92\textwidth]{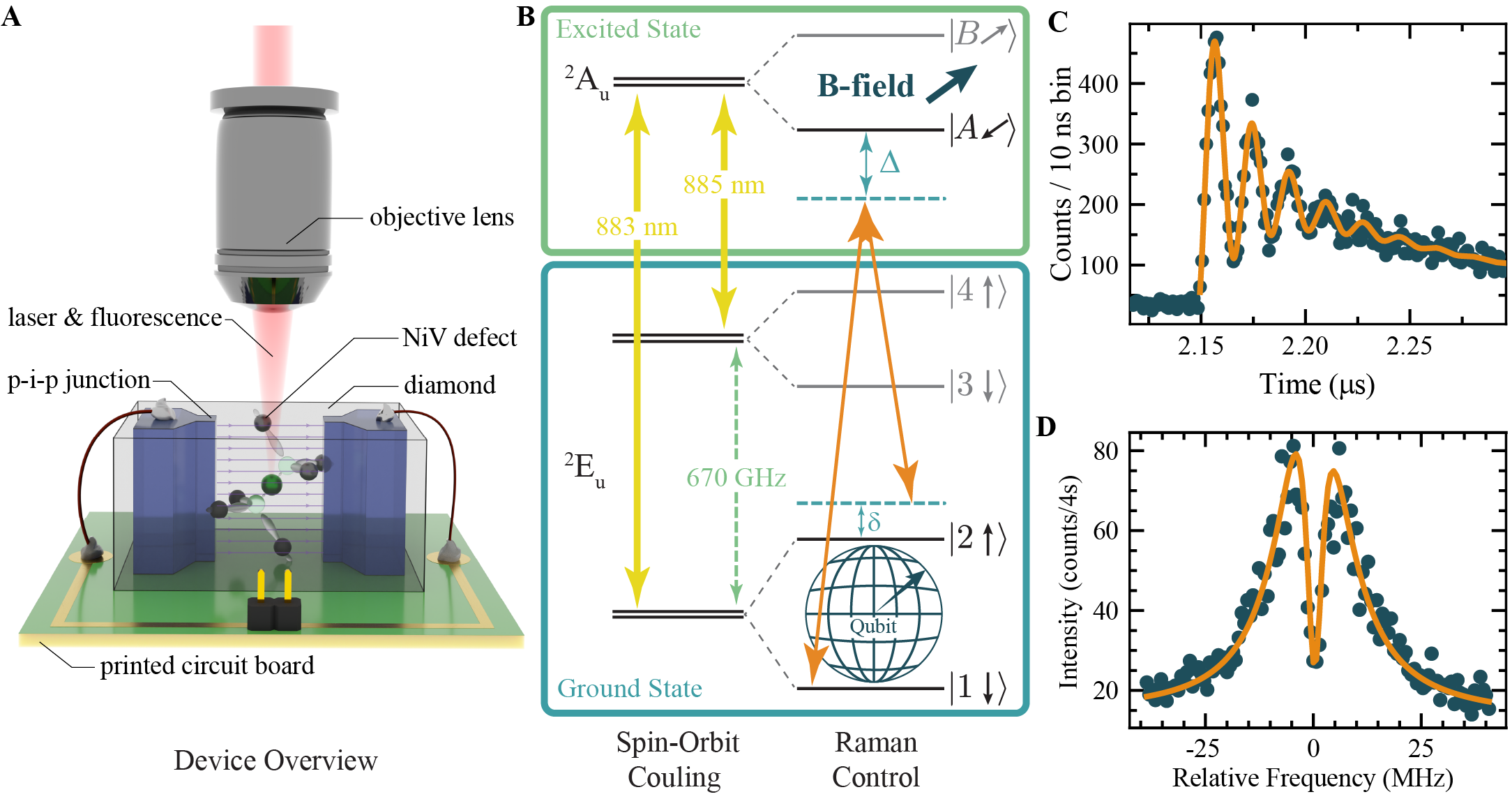}
    \caption{\textbf{An all-diamond quantum device electrically stabilizes a single \NiV spin enabling optical coherent access.} \textbf{A:} Artistic representation of device showing NiV defect inside intrinsic diamond in between two boron doped diamond regions forming a p-i-p junction. The device is electrically contacted via a printed circuit board. Laser excitation and fluorescence collection is achieved via a cryogenic objective lens. \textbf{B:} Electronic structure of the \NiV with spin-orbit and Zeeman interactions shown. Yellow arrows represent resonant optical transitions, orange arrow depict off-resonant optical Raman drive. The magnetic field is applied 54.7$^\circ$ off-axis from (111). \textbf{C:} Optical spin initialization addressing A1 transition with resonant laser pulse. Oscillations indicate coherence between ground and excited state. An initialization fidelity of F=99.1(1)\% is reached after 1 \ensuremath{\mu}s. \textbf{D:} Coherent population trapping with carrier on A1 and EOM sideband on A2. Orange line represents fit with 3-level Hamiltonian model indicating a CPT dip width of 1.32 MHz. All measurements taken at 150 mT and 1.65 K.\\}
    \label{fig:1}
    \end{figure*}

These limitations motivated the search for inversion-symmetric diamond defects with stronger coherent optical emission. The negatively charged silicon-vacancy (\SiV) is the leading example: approximately 70\% of its fluorescence is emitted into the ZPL \cite{ref19} and inversion symmetry suppresses its first-order sensitivity to electric-field noise \cite{ref20,ref21,ref22,ref23}. These properties have enabled integration into nanophotonic devices, enhancing efficient light-matter interactions \cite{ref24,ref25}. However, \SiV replaces the \NV center's optical bottleneck with a spin-coherence bottleneck. Its orbital doublet ground state with relatively small 50GHz spin-orbit splitting leaves phonon-mediated orbital transitions efficient at liquid-helium temperatures, rapidly dephasing the spin \cite{ref23,ref26,ref27}. Millisecond coherence therefore requires either costly and complex millikelvin operation or careful strain engineering to increase the splitting and freeze out these phonon processes \cite{ref28,ref29,ref30}. Nevertheless, these techniques have enabled metropolitan-scale networking with cavity-coupled \SiV qubits \cite{ref31}. A noteworthy alternative is the neutral charge state which however relies on careful Fermi-level engineering, surface preparation, or non equilibrium optical control rather than a broadly reproducible, device-compatible stabilization strategy. Efficient spin polarization and readout additionally require higher-lying bound-exciton transitions \cite{ref32,ref33,ref34}.

This trade-off between optical properties and spin-coherence in \NV and \SiV centers once again motivated the search for defects that preserve favorable optical properties but where phonon protection is built into the electronic structure rather than requiring external engineering. The heavier group-IV vacancy centers, \GeV, \SnV, and \PbV, provide such a built-in route. They retain strong ZPL emission and inversion symmetry, while the heavier impurity atoms produce substantially larger spin-orbit splittings. \cite{ref35,ref36,ref37,ref38}, shifting the relevant phonon transitions to higher frequencies and thus reducing thermal phonon occupation \cite{ref27,ref38}. This enables millisecond spin coherence in \SnV at 1.6K \cite{ref39,ref40,ref41}. This intrinsic protection, however, introduces a new qubit control challenge: microwave transitions between the ground-state qubit levels are suppressed because the states differ in orbital character, whereas strong spin-orbit coupling in the ground- and excited-state manifolds aligns their spin quantization axes, making the optical transitions predominantly spin cycling, even in very strong off-axis magnetic fields. Impressively, both microwave and optical control have been demonstrated for \SnV using strain as an additional symmetry breaking interaction \cite{ref39,ref40,ref41}. Yet strain-assisted control complicates device engineering and can reduce spin lifetimes by increasing the overlap between the qubit states. These constraints reflect a deeper commonality of the group-IV vacancy family. Despite their wide range of spin-orbit splittings, \SiV, \GeV, \SnV, and \PbV share a low-energy structure derived from the same carbon dangling-bond manifold with no significant hybridization from the impurity \emph{p} orbitals. Increasing impurity mass therefore improves phonon protection without fundamentally altering their orbital structure.

Transition-metal impurities provide a different design lever: partially occupied \emph{d} orbitals can hybridize with the carbon dangling bond states, modifying the defect's electronic structure that governs both spin coherence and optical control while preserving the inversion symmetry of the split-vacancy geometry. The negatively charged nickel-vacancy center, \NiV, exemplifies this strategy. Hybridization between nickel \emph{d} orbitals and the split-vacancy carbon states selectively reconstructs the excited-state electronic structure, stabilizing an orbital-singlet absent in the group-IV vacancy family, despite the shared \emph{$D_{3d}$} symmetry, while preserving an orbital doublet ground state with large spin-orbit splitting. This enables direct optical access to the ground-state spin via the excited state singlet, offering a route to long-lived phonon-protected and optically controllable spin qubits without dilution refrigeration or strain engineering. In our previous work, we spectroscopically studied charge stabilized single \NiV centers in electrically contacted all-diamond p-i-p junctions \cite{ref42}. Magneto-optical spectroscopy confirmed the computationally proposed split-vacancy geometry and electronic structure \cite{ref43}, including a large ground-state spin-orbit splitting of approximately 670 GHz and near-infrared and lifetime-limited emission at 883 nm, well suited for direct free-space and short-reach fiber transmission as well as quantum frequency conversion to the telecom C-band. The central question remaining is whether this electronic structure can support the full functionality required of a spin-photon interface. Here, we show that it can. Using actively electrically stabilized \NiV centers in electronic grade (\textless1ppb {[}B{]}, \textless5ppb {[}N{]}) natural $^{13}$C abundance (1.1\%) diamond (c.f. Fig. 1A), we optically initialize, read out, and coherently control the ground-state spin, and use all-optical dynamical decoupling to suppress nuclear-spin-bath-induced dephasing. The resulting coherence exceeds one millisecond at 1.65 K, without dilution refrigeration or strain induced orbital mixing, establishing \NiV as a long-lived, all-optically controlled diamond spin qubit and a proof of principle for engineering spin-photon interfaces through transition-metal orbital hybridization.

\begin{figure}[t]
\centering
\includegraphics[width=0.48\textwidth]{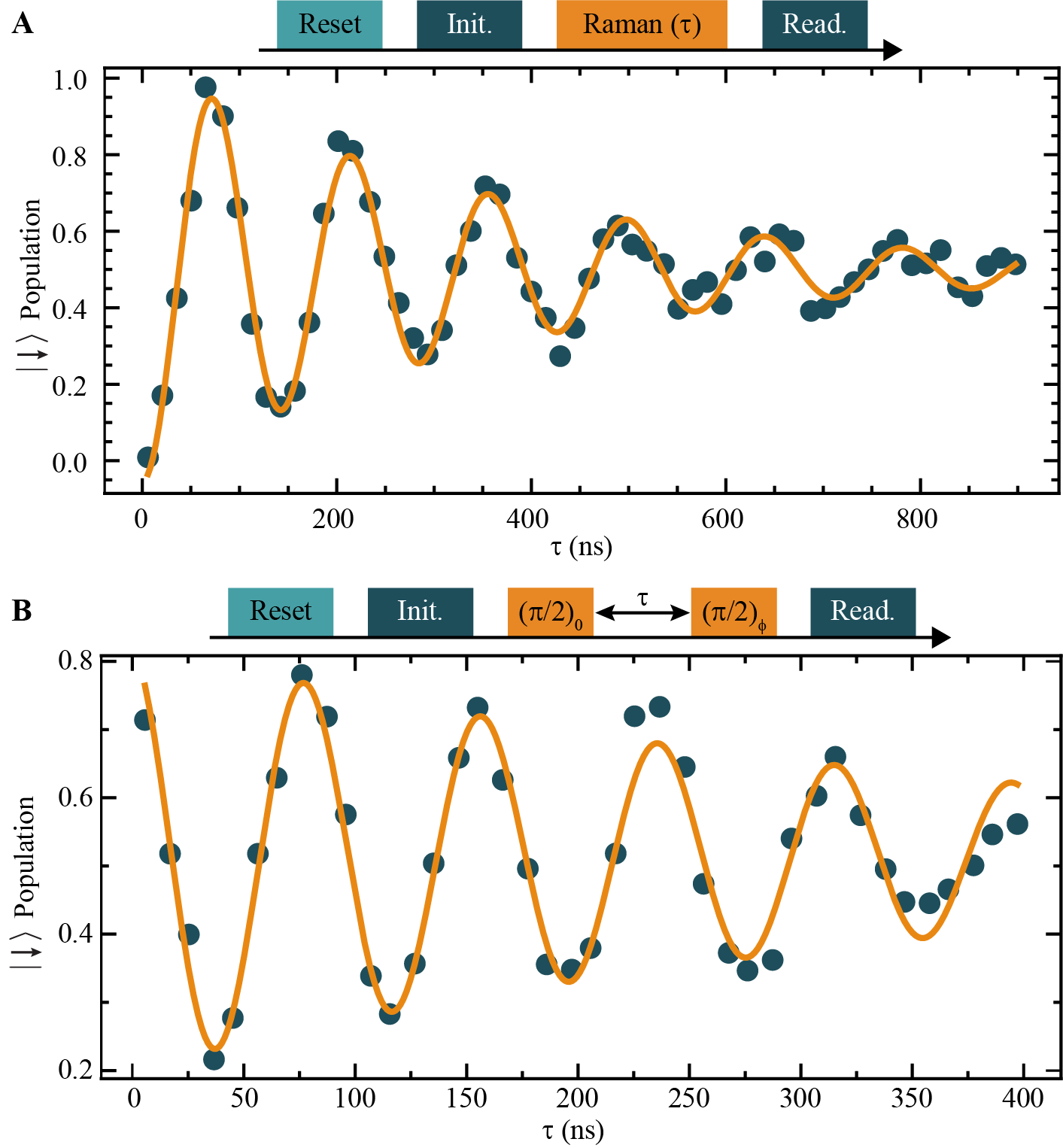}
\caption{\textbf{Raman pulses provide coherent all-optical control of the \NiV ground-state spin. A:} Raman Rabi oscillations between ground state spin levels $\ket{1} = \ket{\downarrow}$ and $\ket{2} = \ket{\uparrow}$. Population in $\ket{\downarrow}$ as a function of optical Raman drive time (blue circles). Fit to a two-level model (orange line). \textbf{B:} Ramsey interferometry with two \ensuremath{\pi}/2 pulses separated by free evolution time \ensuremath{\tau}. The phase of the second \ensuremath{\pi}/2 pulse is rotated as \ensuremath{\phi} = $\tau \omega_s$, with $\omega_s/2\pi$ = 7.5 MHz. Population in $\ket{\downarrow}$ as a function of \ensuremath{\tau} (blue circles) and fit to \(P_{\downarrow} = a\exp\left( -\tau/T_2^* \right)\sin\left( \omega_{Ramsey}\tau + \alpha \right) + b\) (orange line). We extract \(T_{2}^{*} = 371(53)\) ns, \(\alpha = 0.176\), \(\frac{\omega_{Ramsey}}{2\pi} =\)12.593(61) MHz, \(a = 0.313(21)\) and \(b = 0.5143(57)\). All measurements taken at detuning \ensuremath{\Delta}/2\ensuremath{\pi} = 250 MHz, B = 150 mT, and T = 1.65 K.}
\label{fig:2}
\end{figure}

\paragraph*{Optical access to a single \NiV spin.}

The qubit is encoded in the two spin sublevels $\ket{1}$ and $\ket{2}$ of the lower ground-state orbital branch. Under any nonzero off-axis magnetic field, both states can couple to any of the excited state spin sublevels (e.g. $\ket{A}$), forming a $\Lambda$-system (Fig 1B). This direct optical access does not require strain-induced mixing or a threshold magnetic field because the spin quantization axis of the orbital-singlet excited state freely follows the angle of the applied field. We apply a magnetic field of B=150 mT normal to the (100) diamond sample surface, corresponding to a tilt of 54.7$^\circ$ relative to the \NiV symmetry axis, resulting in non-zero $\ket{A} \rightarrow \ket{1}$ (A1) and $\ket{A} \rightarrow \ket{2}$ (A2) optical transitions. Resonant excitation of either transition pumps population into the opposite, undriven ground-state spin level through spontaneous emission. We characterize this process using the branching ratio \(\eta = P_{A1}/P_{A2}\), with transition probabilities \(P_{A1}\) of the stronger, predominantly cycling A1 and \(P_{A2}\) of the weaker nominally spin-flipping A2 transition. Upon application of a magnetic field, we observe immediate optical spin pumping, even at fields as low as 30 mT. Importantly, \(\eta\) is only a function of magnetic field angle not magnitude and we can achieve efficient spin initialization in approx. 200 ns with fidelity $F_{\mathrm{init}}$ \textgreater{} 99\% (Fig. 1C).

To establish coherent optical access to the ground-state spin, we simultaneously drive the A1 and A2 transitions. At two-photon resonance, destructive interference prepares a dark superposition of both ground state spin sublevels, suppressing fluorescence, a phenomenon known as coherent population trapping (CPT) (Fig. 1D). The linewidth of the CPT dip provides a lower bound on the ground-state coherence time, still including potential broadening from optical scattering, power broadening, and technical noise. Extrapolating the power-dependent linewidth to zero power yields $\Delta\nu_0$=0.91(6) MHz (c.f. supplemental material) corresponding to a lower bound of the inhomogeneous dephasing time of $T_2$\textsuperscript{* CPT} = 350(23) ns.

\paragraph*{Coherent qubit control with light.}
Having established a coherent optical \(\Lambda\) system through CPT, we next use it to perform coherent rotations of the ground-state spin. We implement coherent control in a detuned Raman configuration, in which the two optical fields couple the ground state qubit through a virtual excited state. The common single-photon detuning suppresses excited-state population and optical scattering induced decoherence while retaining an effective two-photon coupling between the ground-state spin levels. We use $\Delta/2\pi=250$ MHz and P=55$P_{\mathrm{sat}}$, balancing residual scattering and fast Raman control (see supplemental material for corresponding model).

Following resonant optical reset and initialization, a detuned Raman pulse drives rotations between $\ket{1}$ and $\ket{2}$. The transferred population is read out by again applying a spin-selective resonant laser pulse. We observe coherent Raman Rabi oscillations as a function of Raman-pulse duration with Rabi frequencies of several MHz (Fig. 2A). We next use Ramsey interferometry to establish phase-coherent control of the \NiV ground-state qubit. Two Raman \ensuremath{\pi}/2 pulses separated by a free-evolution time \ensuremath{\tau} map accumulated qubit phase onto spin population, while minimizing optical exposure during the coherence interval. The Ramsey contrast decays with an inhomogeneous coherence time $T_2^*$= 371(53) ns (Fig. 2B), providing a more precise measurement of the center's inhomogeneous coherence time. Sweeping the phase of the second \ensuremath{\pi}/2 pulse produces sinusoidal Ramsey fringes, confirming phase-coherent control of the ground-state qubit and demonstrating that the optical Raman phase controls the qubit rotation axis. The exponential decay of the Ramsey curve reflects residual dephasing which may originate from non-Markovian quasi-static fluctuations, such as magnetic noise from the surrounding $^{13}$C spin bath, or faster relaxation processes, including residual phonon-mediated decoherence. We distinguish these limits using spin refocusing and dynamical decoupling techniques.

\begin{figure}[t]
\centering
\includegraphics[width=0.48\textwidth]{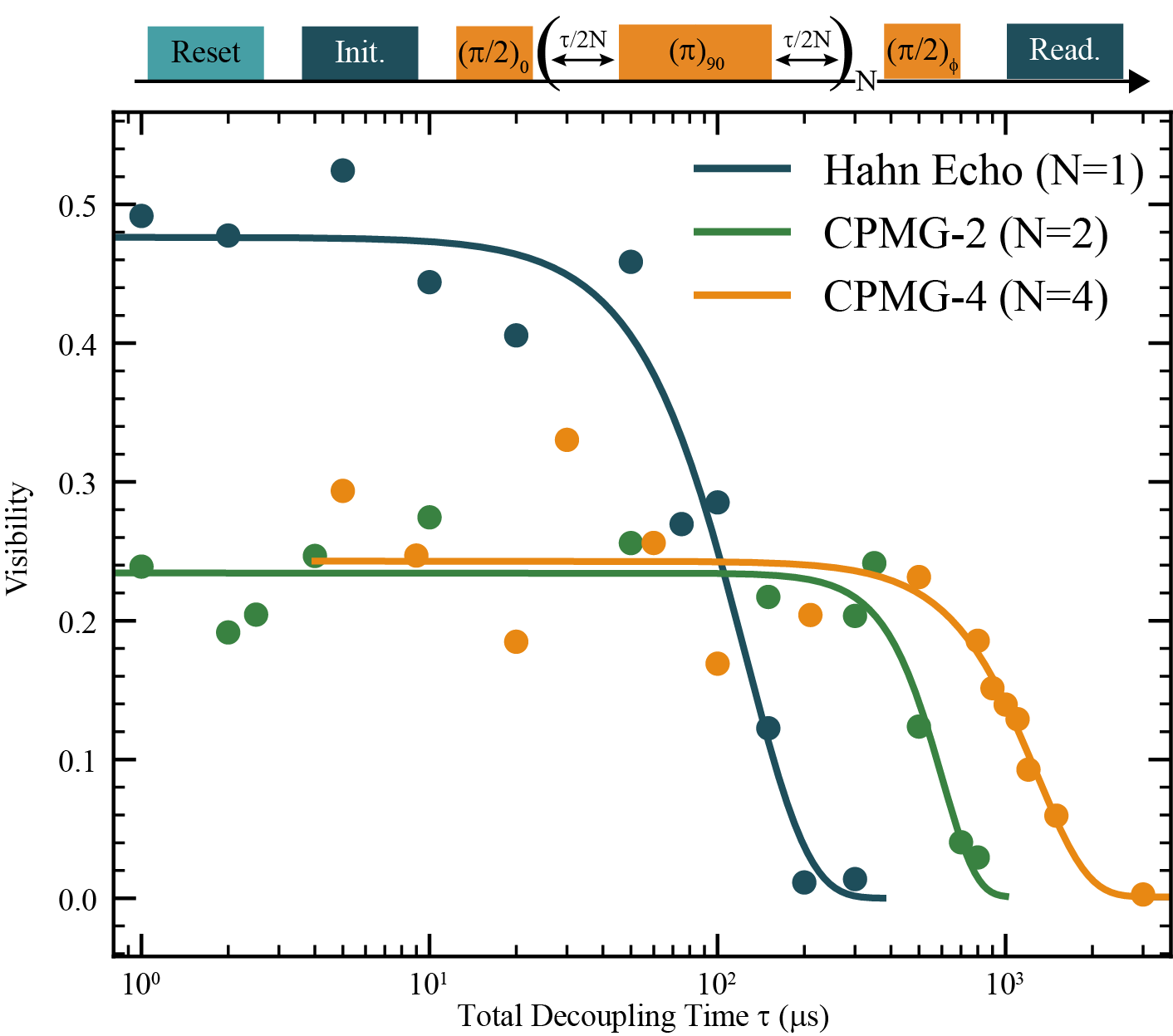}
\caption{\textbf{All-optical dynamical decoupling extends the spin coherence into the millisecond regime.} Hahn Echo and CPMG dynamical decoupling sequences with varying numbers N of refocusing \ensuremath{\pi} pulses. Visibility decay as a function of total decoupling time \ensuremath{\tau} (circles) and fit to stretched exponential \(V(\tau) = v_{0}\exp\left\lbrack - \left( \frac{\tau}{T_{2}} \right)^{n} \right\rbrack + v_{\infty}\) (lines). The Hahn echo fit yielded $V_0$ = 0.476 \ensuremath{\pm} 0.056, $T_2^{\mathrm{echo}}$ = 125 \ensuremath{\pm} 18 \ensuremath{\mu}s, n = 1.99 \ensuremath{\pm} 0.57, and $V_\infty$ = 0.000 \ensuremath{\pm} 0.049, with $R^2$ = 0.958. For CPMG-2, the extracted parameters were $V_0$ = 0.233 \ensuremath{\pm} 0.079, $T_2$ = 601 \ensuremath{\pm} 139 \ensuremath{\mu}s, n = 3.66 \ensuremath{\pm} 1.93, and $V_\infty$ = 0.001 \ensuremath{\pm} 0.076, with $R^2$ = 0.900. For CPMG-4, the fit resulted in $V_0$ = 0.242 \ensuremath{\pm} 0.047, $T_2$ = 1.277 \ensuremath{\pm} 0.229 ms, n = 2.44 \ensuremath{\pm} 1.21, and $V_\infty$ = 0.001 \ensuremath{\pm} 0.044, with $R^2$ = 0.806.The higher absolute visibility of the CPMG-4 dataset reflects improvements to the experimental setup made after the CPMG-2 measurement. See Supplementary Information for more detail. All measurements taken at detuning \ensuremath{\Delta}/2\ensuremath{\pi} = 250 MHz, B=150 mT, and T=1.65 K.}
\label{fig:3}
\end{figure}

\paragraph*{Extending coherence with optical refocusing.}

To determine whether the Ramsey decay is caused by refocusable slow noise or by faster irreversible sources, we next perform all-optical Hahn-echo measurements \cite{ref44}.  The sequence inserts a Raman \ensuremath{\pi} pulse midway between the two \ensuremath{\pi}/2 pulses, reversing phase accumulation from noise that is approximately static over the measurement time. Slow detuning errors and magnetic-field fluctuations can therefore be refocused, whereas rapidly fluctuating or Markovian noise cannot. Figure 3A shows the Hahn-echo visibility as a function of total free-precession time \ensuremath{\tau}. The decay is fit with a stretched exponential
\(v(\tau) = v_{0}\exp\left\lbrack - \left( \frac{\tau}{T_{2}} \right)^{n} \right\rbrack + v_{\infty}\), yielding $T_2^{\mathrm{echo}}$=125(18) \ensuremath{\mu}s. This nearly three-orders of-magnitude increase over the Ramsey value, shows that the dominant dephasing is reversible. The result is consistent with slowly varying magnetic noise, most likely from the surrounding $^{13}$C nuclear spin bath, rather than fast phonon-mediated or Markovian decoherence. The fitted exponent n=2 further supports this interpretation, indicating a noise spectrum dominated by slow, correlated fluctuations on the timescale of the echo sequence \cite{ref45,ref46,ref47}.

To further suppress this low-frequency noise, we implement all-optical Carr-Purcell-Meiboom-Gill (CPMG) sequences with multiple Raman refocusing pulses \cite{ref48,ref49}. Increasing the number of refocusing pulses samples the noise more frequently and extends the filter function to higher frequencies, making CPMG a tool to extend coherence beyond the Hahn-echo limit and a probe of the dominant noise spectrum. Figure 3 shows that CPMG dynamical decoupling substantially extends the \NiV spin coherence. A CPMG-2 sequence increases the coherence time to $T_2^{\mathrm{CPMG-2}}$=601(139) \ensuremath{\mu}s while CPMG-4 extends the coherence into the millisecond regime, yielding $T_2^{\mathrm{CPMG-4}}$=1.28(23) ms. Together with the Hahn-echo result, the continued improvement under additional refocusing pulses shows that the dominant decoherence is low-frequency and refocusable rather than set by fast Markovian relaxation. The fitted stretch exponents remain close to n\ensuremath{\approx}2, consistent with slowly varying magnetic noise, likely from the natural-abundance $^{13}$C nuclear spin bath. Because this noise source can be systematically suppressed through isotopic enrichment with $^{12}$C, these results suggest a clear materials route toward substantially longer spin coherence, potentially approaching the predicted phonon-limited timescale of 30 ms at 1.65 K. The persistence of millisecond-scale coherence under repeated resonant control and readout establishes the all-diamond p-i-p junction as an effective platform for stabilizing the \NiV charge state during coherent quantum operation, eliminating auxiliary optical repump fields that could otherwise introduce charge noise, decoherence, or heating.

    \begin{figure}[t]
    \centering
    \includegraphics[width=0.48\textwidth]{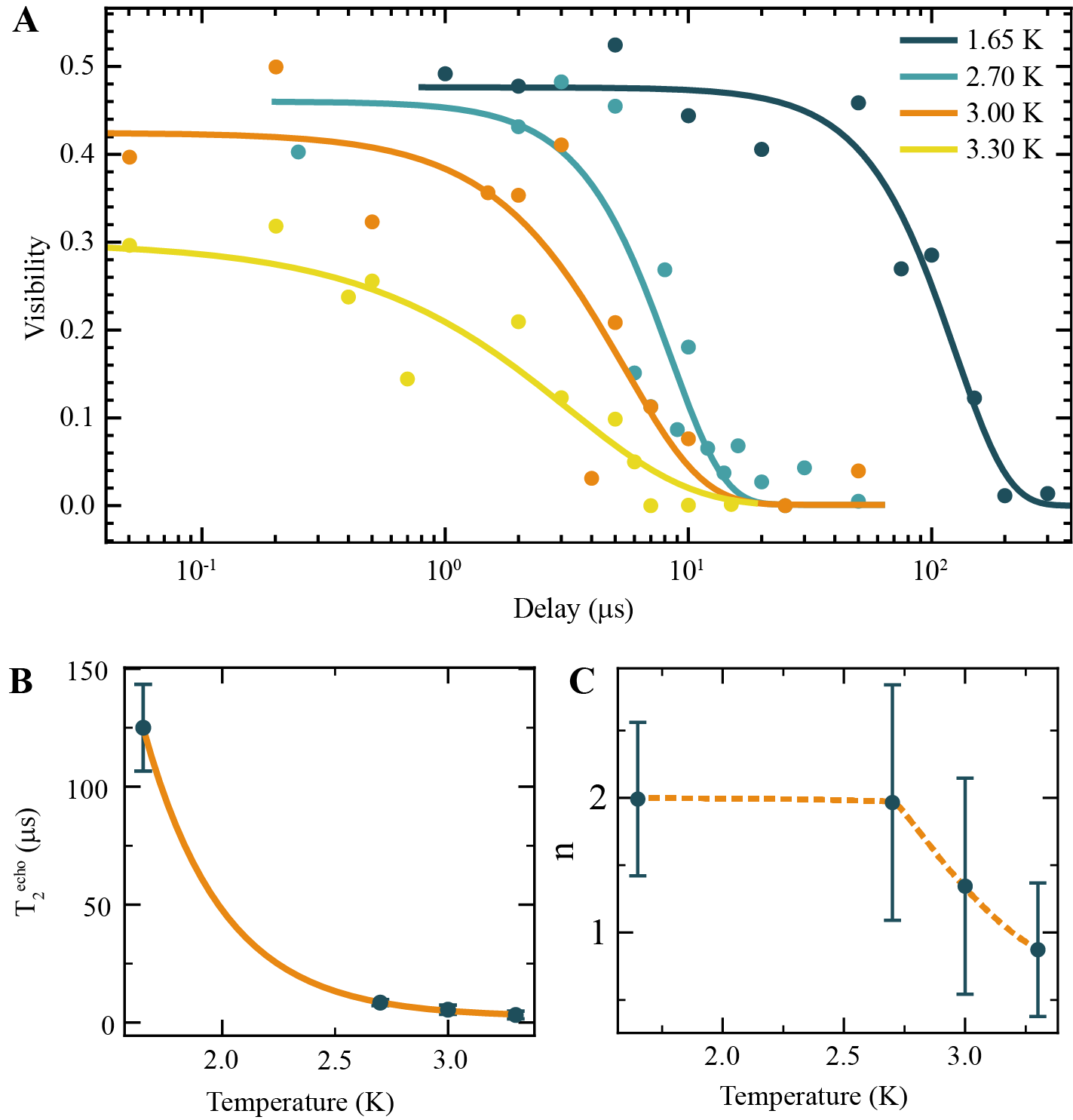}
    \caption{T\textbf{emperature-dependent echo measurements reveal the crossover from spin-bath- to phonon-limited coherence. A:} Hahn echo visibility decay data (dots) for four different temperatures and fits with stretched exponential (lines). \textbf{B:} $T_2^{\mathrm{echo}}$ coherence times as a function of temperature (blue dots) and exponential fit (orange line). We use the 1.65 K measurement from Figure 3 in which $T_2$ = 125 (18) \ensuremath{\mu}s. For 2.7 K, we extract $T_2$ = 8.4 (13) \ensuremath{\mu}s. For 3 K, $T_2$ = 5.5 (19) \ensuremath{\mu}s. For 3.3 K, we extract $T_2$ = 3.2 (17) \ensuremath{\mu}s \textbf{C:} Stretch coefficient n as a function of temperature. Dotted orange line is a guide to the eye. The corresponding stretch exponents are: n=1.99(57) at 1.65 K, n=1.97(87) at 2.7 K, n=1.34(80) at 3.0 K, and n=0.87(50) at 3.3 K.}
    \label{fig:4}
    \end{figure}

\paragraph*{Temperature limits to quantum memory.}

We investigate the temperature dependence of the spin coherence to gain further insights into the transition from a spin-bath limited to a phonon limited regime. Figure 4A shows full Hahn-echo visibility decays measured at four temperatures. As temperature increases, the coherence time and the stretch exponent decrease. The extracted coherence times are summarized in Fig. 4B and are fit using an exponential temperature dependence assuming resonant one-phonon absorption \cite{ref27} inducing transitions between orbital branches and randomizing the spin state phase. Figure 4C shows the corresponding stretch exponents. At 3.0 and 3.3 K, the fitted values have visibly decreased, approaching n=1, consistent with a fast memory-less Markovian phonon bath. These measurements identify 1.65 K as a regime in which \NiV coherence remains dominated by refocusable spin-bath noise rather than phonons, placing millisecond quantum memory within reach of compact cryogen-free closed-cycle cryostats rather than dilution refrigeration. For short-reach links, such as data-center, campus, or metropolitan-scale deployments where the required storage time can be substantially shorter, operation in even more compact 2.3 K cryogenic platforms may be sufficient.

\paragraph*{Engineered transition-metal qubits in diamond.}

The emergence of millisecond spin coherence under all-optical control shows that the \NiV center is not merely a spectroscopically attractive model system, but a highly functional spin-photon interface whose properties follow from well-defined electronic-structure design rules. Combining these properties with isotopically purified diamond, nanophotonic enhancement, and telecom frequency conversion could therefore yield compact, optically controlled quantum-network nodes based on defects whose coherence and optical interface are built into the same microscopic system.

More broadly, these results establish electronically and optically active transition-metal impurities as a new design space for diamond spin-photon interfaces. \NiV shows that \emph{d}-orbital hybridization can reshape the optical manifold while preserving the inversion symmetry and spin-orbit-protected ground-state structure of the split-vacancy geometry. In future systems, such hybridization could even yield an inversion-symmetric defect with a spinful orbital-singlet ground state, combining the robust spin properties of \NV-like qubits with the optical stability of split-vacancy centers. The broader chemical space of transition-metal impurities therefore offers a path toward diamond defects with tailored combinations of long coherence, stable optical emission, and direct optical controllability.\\

\begin{acknowledgments}
\paragraph*{Funding}
National Science Foundation grant PHY-2608129 (JNB)\\
Cowen Family Endowment (JNB)\\
MSU Alfred and Ruth Zeits Fellowship (IMM)\\
MSU Venture Fellows Postdoctoral Program (IMM)\\

\paragraph*{Author contributions}
\begin{quote}
Conceptualization: JNB, IMM, TA\\
Methodology: JNB, IMM, TA, LC, SP, JM\\
Investigation: JNB, IMM, TA, TL\\
Visualization: IMM, JNB\\
Funding acquisition: JNB, SSN, SP, JM\\
Project administration: JNB\\
Supervision: JNB, SSN, SP, JM\\
Writing -- original draft: IMM, JNB\\
Writing -- review \& editing: IMM, JNB, TA, SSN, JM, SP, DJT\\
\end{quote}

\paragraph*{Competing interests} Authors declare that they have no
competing interests.\\

\textbf{\hfill\break
}
\end{acknowledgments}

\clearpage
\onecolumngrid
\setcounter{section}{0}
\setcounter{subsection}{0}
\setcounter{figure}{0}
\setcounter{table}{0}
\setcounter{equation}{0}
\renewcommand{\thefigure}{S\arabic{figure}}
\renewcommand{\thetable}{S\arabic{table}}
\renewcommand{\theequation}{S\arabic{equation}}
\renewcommand{\thesection}{S\arabic{section}}
\renewcommand{\thesubsection}{\thesection.\arabic{subsection}}
\renewcommand{\theHfigure}{supp.\arabic{figure}}
\renewcommand{\theHtable}{supp.\arabic{table}}
\renewcommand{\theHequation}{supp.\arabic{equation}}
\renewcommand{\theHsection}{supp.\arabic{section}}

\begin{center}
{\large\bf Supplemental Material}\\[0.75em]
{\bf A transition-metal qubit in diamond with all-optical control and millisecond quantum memory}
\end{center}

\section{Materials and Methods}

\subsection{Sample and cryogenic optical setup}

Measurements were performed on a single negatively charged nickel-vacancy center, \NiV, in an electrically contacted all-diamond p-i-p junction device. The device architecture and charge-stabilization procedure are described in \cite{ref42}.
The built-in electrical charge control stabilized the \NiV charge state under resonant excitation, eliminating the need for an additional optical repump beam during the spin-control measurements. For all measurements a bias of 20 V was applied to the junction.

The sample was mounted in a closed-cycle helium cryostat (Attocube attoDRY2100) and cooled to 1.65 K. Optical excitation and fluorescence collection were performed using a confocal microscope integrated with the cryostat, consisting of a nanopositioning system (attocube ANPx101/RES/LT, ANPz102/RES/LT, ANSxy100std/LT, ANSz100std/LT) and a cryogenic apochromatic objective lens (attocube LT-APO/VISIR/0.82) . A superconducting solenoid integrated into the cryostat was used to apply a magnetic field in Faraday configuration normal to the (100)-oriented
diamond surface. For this geometry, the applied field is tilted by 54.7$^\circ$ relative to the \NiV symmetry axis. This off-axis field configuration produces finite overlap between the spin eigenstates of the orbital-singlet excited state and both ground-state spin projections, enabling optical access to the otherwise spin-forbidden transition. Unless otherwise stated, coherent-control and coherence measurements were performed at B=150 mT.

\subsection{Optical tone generation and pulse control}

Multiple optical frequencies were generated using an electro-optic phase modulator (EOM; EOSpace) driven by an arbitrary waveform generator (Tektronix AWG70001A). For spectroscopy and coherent population trapping, the optical carrier was parked on one transition and radio-frequency modulation of the EOM generated sidebands used to address the second transition. 

For resonant reset, initialization, and readout pulses, optical pulses were generated using serrodyne-modulated waveforms applied to the EOM. This allowed the relevant resonant sideband frequency to be pulsed on and off while maintaining a fixed off-resonant weak optical carrier. For detuned Raman control, sinusoidal modulation of the EOM generated two first-order sidebands whose frequency separation matched the ground-state two-photon transition. The optical carrier was offset from the midpoint between the (A1) and (A2) transitions so that both Raman fields were detuned from the excited state by the same single-photon detuning. Unless otherwise stated, Raman-control measurements used a single-photon detuning of 250 MHz. The resonant optical power used for the pulsed spin measurements was 14 \ensuremath{\mu}W, measured in front of the objective lens (c.f. model of Raman control below). An additional acousto-optical modulator (AOM; Isomet) was used to completely switch off the optical fields during free evolution periods to rule out any unwanted optical pumping effects.

\subsection{Three-Level Model for Coherent Population Trapping}
The coherent population trapping spectrum was modeled using a three-level $\Lambda$ system consisting of two ground state spin levels, $\ket{\downarrow}$ and $\ket{\uparrow}$, coupled optically to a common excited state, $\ket{A}$. The two optical fields drive the $\ket{\downarrow} \leftrightarrow \ket{A}$ and $\ket{\uparrow} \leftrightarrow \ket{A}$ transitions with Rabi frequencies \ensuremath{\Omega}\ensuremath{_1} and
\ensuremath{\Omega}\ensuremath{_2}, respectively. The steady-state excited-state population was
calculated by solving a Lindblad master equation at each value of the two-photon detuning. In the rotating-wave approximation, the Hamiltonian
in the basis $\left\{\ket{\downarrow}, \ket{\uparrow}, \ket{A}\right\}$ is \cite{ref39, ref50}
\[H\  = \begin{pmatrix}
0 & 0 & \frac{\Omega_1}{2} \\
0 & - \delta & \frac{\Omega_2}{2} \\
\frac{\Omega_1}{2} & \frac{\Omega_2}{2} & \Delta
\end{pmatrix}\]

Here, \ensuremath{\delta} is the two-photon detuning, defined as the difference between the frequency separation of the two optical fields and the energy splitting between the two ground states, while \ensuremath{\Delta} is the common single-photon detuning of both optical fields from their respective optical transitions to the excited state. The diagonal term -\ensuremath{\delta} assigned to |\ensuremath{\uparrow}\ensuremath{\rangle}
represents the relative energy mismatch between the two ground states in
the rotating frame. When \ensuremath{\delta} = 0, the system is on two-photon resonance
and can evolve into a dark-state superposition that is decoupled from
the excited state. This produces the narrow reduction in fluorescence
characteristic of coherent population trapping. To account for
irreversible decoherence mechanisms, we use a Lindblad Master equation.
The density matrix \ensuremath{\rho} evolves according to \cite{ref39, ref50}
\[d\rho/dt\  = \  - i\lbrack H,\rho\rbrack\  + \ \sum_{k}^{}{\ \lbrack C_{k}\rho C_{k}^{\dagger}\  - \ \frac{1}{2}(C_{k}^{\dagger}C_{k}\rho + \rho C_{k}^{\dagger}C_{k})\rbrack}.\]

The steady-state solution is obtained by imposing $d\rho/dt\  = \ 0$ together with the normalization condition
$Tr(\rho)\  = \ 1$. The model includes both coherent optical evolution through H
and incoherent decay and dephasing through the collapse operators C\ensuremath{_j}.
Spontaneous decay from the excited state to the two ground states is
represented by
\[
C_1=\sqrt{\gamma_{\downarrow,A}}\,\ket{\downarrow}\bra{A},
\qquad
C_2=\sqrt{\gamma_{\uparrow,A}}\,\ket{\uparrow}\bra{A}.
\]

The dominant optical decay rate is fixed by the measured excited-state
lifetime,
\[\gamma_{\downarrow ,A}\  = \ 1/T_{\mathrm{opt}},\]

with $T_{\mathrm{opt}}$ = 10.5 ns in the simulation. The second optical
decay rate is defined using the branching ratio \ensuremath{\eta},
\[\gamma_{\uparrow ,A}\  = \ \gamma_{\downarrow ,A}/\eta.\]

The branching ratio therefore controls the relative probability that excitation followed by spontaneous emission returns the system to
$\ket{\downarrow}$ or transfers it to $\ket{\uparrow}$. Population relaxation between the two ground-state spin levels is modeled using two symmetric collapse operators,
\[
C_3=\sqrt{\gamma_1}\,\ket{\uparrow}\bra{\downarrow},
\qquad
C_4=\sqrt{\gamma_1}\,\ket{\downarrow}\bra{\uparrow},
\]
where the ground-state spin-relaxation rate is $\gamma_1=\frac{1}{T_1}$. These terms repopulate the two spin states and prevent the system from remaining indefinitely in a perfect dark state. Loss of coherence between $\ket{\downarrow}$ and $\ket{\uparrow}$ is included through the operator
\[
C_{\phi}
=
\sqrt{\frac{\gamma_2}{2}}
\left(
\ket{\downarrow}\bra{\downarrow}
-
\ket{\uparrow}\bra{\uparrow}
\right).
\]

The pure-dephasing rate is $\gamma_2=\frac{1}{T_{2,\mathrm{pure}}}$.
This collapse operator suppresses the off-diagonal ground-state coherence $\rho_{\downarrow\uparrow}$ without directly transferring population between the two ground states. Increasing $\gamma_2$ broadens and reduces the depth of the CPT feature. A horizontal dip-shift parameter, $\delta_{\mathrm{shift}}$, is included to account for an off-center resonance. Because the steady-state excited-state population is proportional to the measured optical signal, the excited-state population, $\rho_{AA}$, is converted to the experimentally measured fluorescence according to
\[I(\delta)=\alpha \rho_{\mathrm{AA}}(\delta)+\beta+m\delta.\]

Here, $\alpha$ is an overall scale factor, $\beta$ is a constant count-rate offset,
and m is a linear background slope. The scale factor accounts for the data not being normalized to the excited state population. The offset accounts for
background fluorescence and detector dark counts, while the slope
captures a slowly varying background across the frequency scan. The
optical lifetime and branching ratio are fixed independently. For the
simulation shown in the main text Figure 1, the parameters are: $\alpha$ =
4600, $\beta$ = 13.26, m = -0.01, Delta = 0 MHz, shift = -0.5 MHz,
$\Omega_{1}/2\pi = 10.01$ MHz, $\Omega_{2}/2\pi = 1.01$ MHz,
T\textsubscript{1} = 15 ms, and $T_{2,\mathrm{pure}}$ = 240.38 ns.

\subsection{Power Dependent CPT Measurements}
To extract the dephasing rate from the CPT measurements, we measured CPT as a function of optical power and extrapolated the fitted decoherence rate to the zero-power limit. This data was fit to the three-level model discussed above. The extracted dephasing rates, $\gamma_2$ are shown as a function of optical power in Supplementary
Fig. S1. The optical power was measured before the objective and
multiplied by 0.67 to account for the power in the unused negative-order
modulation sideband. A linear fit to the power-dependent linewidths
yielded a zero-power intercept of 0.91 \ensuremath{\pm} 0.06 MHz. Using
$T_2^{\mathrm{CPT}} = 1/(\pi\gamma_2)$, this corresponds to
$T_2^{\mathrm{CPT}} = 350 \pm 23~\mathrm{ns}$. 

\begin{figure}[t]
\centering
\includegraphics[width=0.5\textwidth]{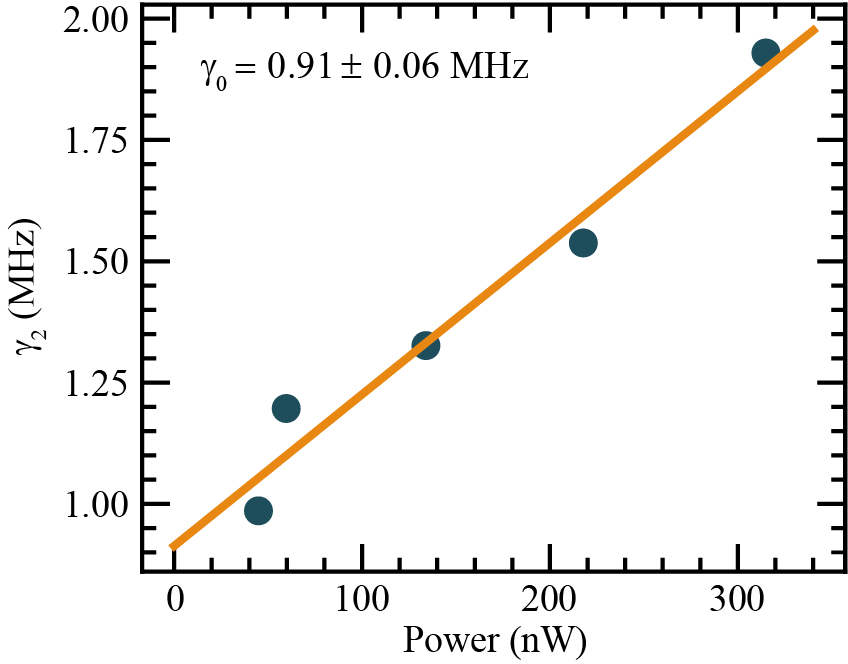}
\caption{Extracted ground state decoherence rate from the
three-level Hamiltonian model for CPT data at different optical powers. A linear
fit is applied finding a y-intercept shown in the figure and a slope of
0.0031(3) MHz/nW with R\ensuremath{^2} = 0.96.}
\label{fig:s1}
\end{figure}

\subsection{Branching Ratio}

The optical branching ratio, \ensuremath{\eta}, was calculated from the
magnetic-field-dependent eigenstates of the \NiV
ground and excited states based on a group-theoretical model originally
developed for the SiV center
\cite{ref23}
and adapted for the NiV center
\cite{ref42}.
The ground-state Hamiltonian includes the spin--orbit interaction, spin
and orbital Zeeman interactions, and any strain or Jahn--Teller terms.
For each magnetic-field magnitude and orientation, the Hamiltonian was
diagonalized to obtain the two lowest energy ground-state eigenstates $\ket{1}$ and $\ket{2}$. The excited state was treated as an orbital singlet
with spin 1/2, and its Zeeman Hamiltonian was diagonalized to obtain the
lower energy excited-state eigenstate $\ket{A}$. The optical transition strengths were evaluated using the orbital dipole operators P\ensuremath{_x}, P\textsubscript{y}, and
P\textsubscript{z} adopted from the established SiV model
\cite{ref23}.
For each ground-state branch, the transition strength was calculated by
summing the squared dipole matrix elements
\[P_{Ai}=\left|\bra{i}P_{x}\ket{A}\right|^2
+
\left|\bra{i}P_{y}\ket{A}\right|^2
+
\left|\bra{i}P_{z}\ket{A}\right|^2.
\]

where $i$ = 1, 2 labels the two ground-state eigenstates. The branching
ratio was then computed as \ensuremath{\eta} = $P_{A1}/ P_{A2}$. Thus, \ensuremath{\eta} represents the ratio of the calculated oscillator strengths of the A1 and A2 transitions. For the magnetic-field orientation used here, in which the field is applied at an angle of approximately 54.7$^\circ$ relative to the defect axis, a branching ratio on the order of \ensuremath{\eta} \ensuremath{\approx} 3.7 was predicted from the group theory model.

\begin{figure}[t]
\centering
\includegraphics[width=0.8\textwidth]{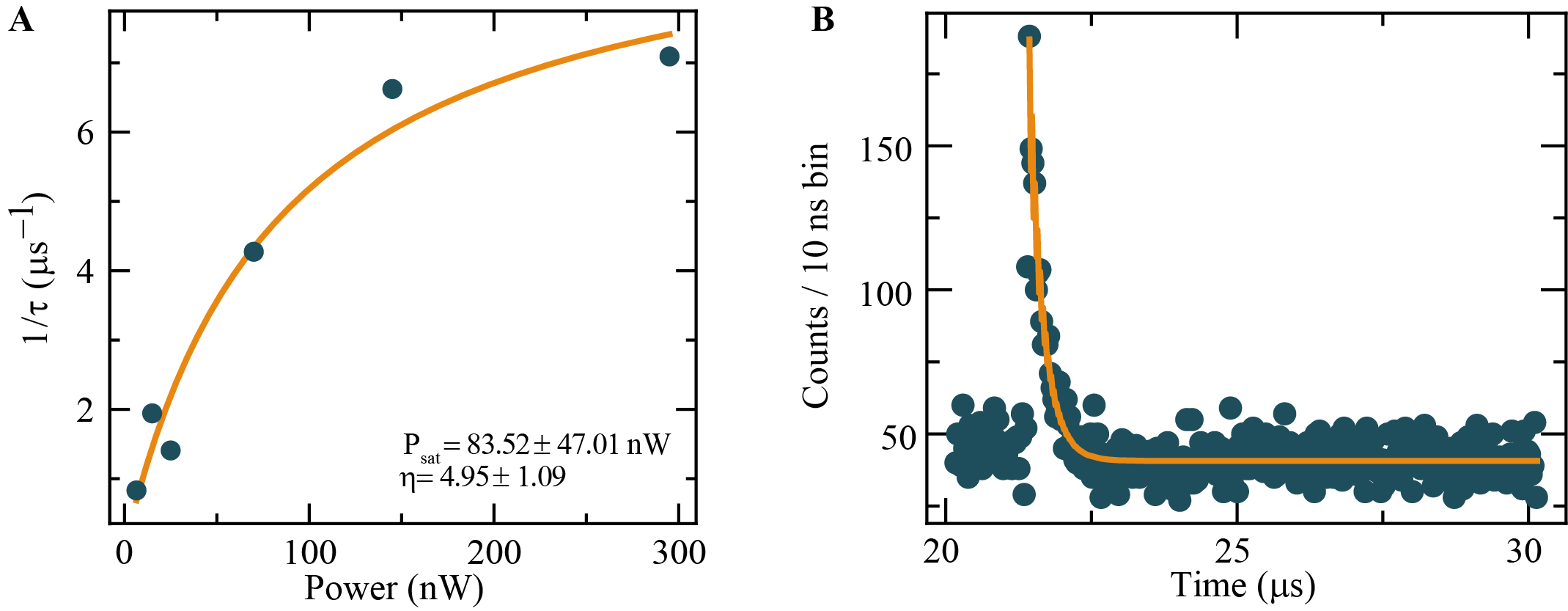}
\caption{\textbf{A} The spin pumping rate as a function of optical
power. This was fit to the equation $\Gamma_{\mathrm{pump}}(p)
=
\frac{\Gamma}{2}
\left(
\frac{p/p_{\mathrm{sat}}}{1+p/p_{\mathrm{sat}}}
\right)
\frac{1}{\eta}$.
to extract \(p_{\mathrm{sat}}\) and \(\eta\) shown in the graph. Each data point
corresponds to an inverted spin pumping time extracted from a spin
pumping measurement consisting of a 10 \ensuremath{\mu}s reset (A2) pulse followed by a
10 \ensuremath{\mu}s initialization (A1) pulse. The spin pumping time extracted was
from the decay associated with pumping A1, which should provide an upper
estimate of \(\eta\). \textbf{B} A representative spin pumping
measurement is shown displaying the fluorescence peak and subsequent
decay associated with the initialization pulse. The power used for this
measurement was 145 nW and the data was fit to an exponential decay curve.}
\label{fig:s2}
\end{figure}

The branching ratio was also determined experimentally from power-dependent spin-pumping measurements. The measured spin-pumping rate depends on both the optical scattering rate and the probability that an individual scattering event changes the spin state. The optical scattering rate follows the saturation behavior of a driven two-level transition, while the spin-pumping rate is reduced by the branching
ratio. Using the approximation that one out of every \ensuremath{\eta} scattering events changes the spin state, the pumping rate is \cite{ref50}
\[
\Gamma_{\mathrm{pump}}(p)
=
\frac{\Gamma}{2}
\left(
\frac{p/p_{\mathrm{sat}}}{1+p/p_{\mathrm{sat}}}
\right)
\frac{1}{\eta}.
\]

where \ensuremath{\Gamma}/2\ensuremath{\pi} = 15 MHz is the \NiV excited-state decay
rate, $p$ is the applied resonant optical power, $p_{\mathrm{sat}}$ is the saturation power, and \ensuremath{\eta} is the branching ratio \cite{ref39}. The corresponding spin-pumping time is $\tau_{\mathrm{pump}}(p) = 1/\Gamma_{\mathrm{pump}}(p)$. At low optical power, \ensuremath{\Gamma}\textsubscript{pump} increases approximately linearly with p. At high power, the optical transition saturates and the pumping rate approaches \ensuremath{\Gamma}/(2\ensuremath{\eta}). Fitting the measured spin-pumping time as a function of optical power therefore provides estimates of both $p_{\mathrm{sat}}$ and \ensuremath{\eta}. A characteristic spin-pumping decay is shown in Figure S2B and the extracted inverted decay times are shown in Figure S2A. The fit gives $p_{\mathrm{sat}}$ = 83.52(47.01) nW and \ensuremath{\eta}=4.95(1.09), as shown in Fig. S2A. The measured bulk value is consistent with the branching ratio expected from the group-theoretical model for the applied magnetic-field magnitude and orientation. The spin pumping time was extracted using a pulse sequence consisting of a 10 $\mu$s reset pulse, followed by a 500 ns break and then a 10 $\mu$s initialization pulse. This was then performed at different powers and the initialization decay time was extracted using an exponential decay.

\subsection{Spin Initialization Fidelity}
Spin initialization fidelity was characterized using a two-pulse sequence consisting of a 1 \ensuremath{\mu}s reset pulse resonant with the A2 transition, followed by a 500 ns laser-off interval and a 1 \ensuremath{\mu}s initialization pulse resonant with the A1 transition. An optical power of 14 \ensuremath{\mu}W, measured before the objective, was used for both pulses. The time-resolved fluorescence was fit using an exponential decay combined with a damped sinusoid to account for coherent optical Rabi oscillations between the ground and excited states during spin pumping, shown in main text Fig. 1B. In particular, the equation was $y(t)=y_{\infty}+A_{\mathrm{pump}}e^{-t/\tau_{\mathrm{pump}}}+A_{\mathrm{osc}}e^{-t/T_{\mathrm{osc}}}\cos\left(2\pi f_{\mathrm{Rabi}}t+\phi\right)$. For the initialization pulse, the extracted optical pumping time was \ensuremath{\tau}\textsubscript{pump} = 106.78 \ensuremath{\pm} 1.24 ns, the Rabi frequency was $f_{\mathrm{Rabi}} = $ 56.131 \ensuremath{\pm} 0.162 MHz, and the oscillation decay time was T\textsubscript{osc} = 25.56 \ensuremath{\pm} 0.68 ns. After subtracting dark counts and background measured during the laser-off interval, an initialization fidelity of $F_{\mathrm{init}}=99.1$ \ensuremath{\pm} 0.1\% was obtained relative to the fluorescence baseline reached after 1\ensuremath{\mu}s. The fidelity was calculated as F = 1 - I\textsubscript{baseline}/I\textsubscript{peak}.

The same analysis was performed for the reset pulse. The extracted optical pumping time was \ensuremath{\tau}\textsubscript{pump} = 99.98 \ensuremath{\pm} 1.12 ns, the Rabi frequency was 49.575 \ensuremath{\pm} 0.150 MHz, and the oscillation decay time was T\textsubscript{osc} =
25.88 \ensuremath{\pm} 0.63 ns. The residual pumped-state population was estimated to be 1.71\%, corresponding to a reset fidelity of F\textsubscript{reset}=98.29\%. The reset pulse is slightly faster at pumping compared to the initialization likely because once excited out of the previously spin forbidden state, it is more likely to decay into the nominally spin cycling $\ket{1}$ state. The opposite is true when pumping out the $\ket{1}$ state, and as such, the time is slightly longer. 

The difference between the initialization and reset fidelities may arise from the frequencies used to generate the two optical sidebands. Because the optical carrier is positioned 250 MHz from the midpoint of the A1 and A2 transitions and closer to A2, the reset pulse is generated using the lower-frequency sideband. Residual background fluorescence from the sample or from a spectrally nearby NiV center may also increase the measured baseline and therefore would not represent a fundamental limit on spin initialization. Alternatively, off-resonant excitation by the carrier or a higher-order modulation sideband may prevent the fluorescence from reaching a lower baseline and limit the measured fidelity.

\subsection{Raman Control Theory}
To mitigate optical-scattering induced decoherence, we move to the far-detuned regime in which the single photon detuning satisfies the following condition \(\Delta \gg \sqrt{s\ \Gamma}\). Here s = P/$P_{\mathrm{sat}}$ is the optical saturation parameter, P is the optical power in each Raman field, $P_{\mathrm{sat}}$ is the measured saturation power, and \ensuremath{\Gamma} is the excited-state decay rate. For the \NiV, \ensuremath{\Gamma}/2\ensuremath{\pi} = 15 MHz. In this limit, the approximate optical scattering rate is \cite{ref39, ref50}
\[\Gamma_{\mathrm{os}}\  = \ (P/P_{sat})(\Gamma ^{3}/8\Delta \ensuremath{^2}).\]

The corresponding optical-scattering-limited relaxation and coherence
times are approximately
\[
T_{1,\mathrm{os}}=\frac{\eta}{\Gamma_{\mathrm{os}}},
\qquad
T_{2,\mathrm{os}}=\frac{1}{\Gamma_{\mathrm{os}}},
\]

where \ensuremath{\eta} is the branching ratio \cite{ref39}. The detuning must therefore balance reduced scattering against sufficiently rapid coherent control. For the measurements reported here, \ensuremath{\Delta}/2\ensuremath{\pi} = 250 MHz and s \ensuremath{\approx} 55. The single-photon detuning and saturation parameter were selected by estimating the fidelity of a Raman-based \ensuremath{\pi}/2 rotation while accounting for both optical-scattering-induced decoherence and inhomogeneous dephasing from $^{13}$C nuclear spin bath. For these calculations, the branching ratio was \ensuremath{\eta} = 5, and the inhomogeneous coherence time was $T_2^*$ = 0.5\ensuremath{\mu}s. Increasing s raises the scattering rate linearly, whereas increasing the detuning suppresses scattering as $1/\Delta^2$. The total dephasing rate was approximated as
\[\Gamma_{\mathrm{tot}}\  = \max(\Gamma_{\mathrm{os}},\ 1/T_{2}^{*}).\]

This expression assumes that the relevant gate decoherence is limited by whichever process is faster: optical scattering or the background inhomogeneous dephasing. Assuming approximately equal optical power in either branch exciting A1 and A2, the Raman-based Rabi frequency was calculated as \cite{ref50}
\[\Omega\  = \ s\Gamma \ensuremath{^2}\ /\ (4\sqrt{\eta}\ \Delta).\]

This expression assumes that \ensuremath{\Gamma}, \ensuremath{\Delta}, and \ensuremath{\Omega} are treated consistently as angular frequencies. Defining a \ensuremath{\pi}/2 gate quality factor as
\[Q_{\pi/2}\  = \ 2\Omega\ /\ (\pi\Gamma_{\mathrm{tot}}).\]

Then, we have
\[1/Q_{\pi/2}\  = \ \Gamma_{\mathrm{tot}}\ t_{\pi/2}.\]

Thus, 1/Q\textsubscript{\ensuremath{\pi}/2} represents the amount of decoherence
accumulated during the \ensuremath{\pi}/2 pulse. Assuming that the coherent
contribution to the state decays exponentially during the gate, the
remaining coherence is
\[C_{\pi/2}\  = \exp( - \Gamma_{\mathrm{tot}}\ t_{\frac{\pi}{2}}) = \exp( - 1/Q_{\pi/2}).\]

The estimated \ensuremath{\pi}/2 gate fidelity is then
\[F_{\pi/2}\  = \ \frac{1}{2}\lbrack 1 + C_{\frac{\pi}{2}}\rbrack = \ \frac{1}{2}\lbrack 1 + \exp( - 1/Q_{\pi/2})\rbrack.\]

This expression maps the surviving coherence onto a fidelity between 0.5 and 1. A value of 1 corresponds to an ideal \ensuremath{\pi}/2 rotation with negligible decoherence, while the fidelity approaches 0.5 when the coherence is completely lost during the pulse. The single photon detuning and saturation parameter were swept to create a 2D fidelity plot as shown in Figure S3. Increasing s increases the Raman Rabi frequency and shortens the \ensuremath{\pi}/2 pulse, but it also increases optical scattering. Increasing \ensuremath{\Delta} suppresses optical scattering more rapidly than it reduces the Raman Rabi frequency, but a larger detuning requires a larger optical power, and therefore a larger s, to maintain a sufficiently fast gate. The selected operating point was chosen from the region where the calculated \ensuremath{\pi}/2 fidelity was high while keeping both the saturation parameter and single-photon detuning experimentally practical.

For the experimental parameters of s \ensuremath{\approx} 55, and \ensuremath{\Delta}/2\ensuremath{\pi} = 250 MHz, the model predicts a Raman Rabi frequency of approximately \ensuremath{\Omega}/2\ensuremath{\pi} = 6.3 MHz. The same model also gives optical-scattering-limited times of T\textsubscript{2,os} \ensuremath{\approx} 0.421 \ensuremath{\mu}s and T\textsubscript{1,os} \ensuremath{\approx} 1.684 \ensuremath{\mu}s.

Using the theoretically predicted values as an initial guess, the Rabi data was then fit to a driven two-level master-equation model. The two basis states \(|0\rangle\) and \(|1\rangle\) represent the lower and upper ground-state spin states coupled by the effective two-photon Raman transition. After adiabatically eliminating the optically excited state, the coherent dynamics were described by the effective Hamiltonian \cite{ref39, ref50}

\begin{figure}[t]
\centering
\includegraphics[width=0.65\textwidth]{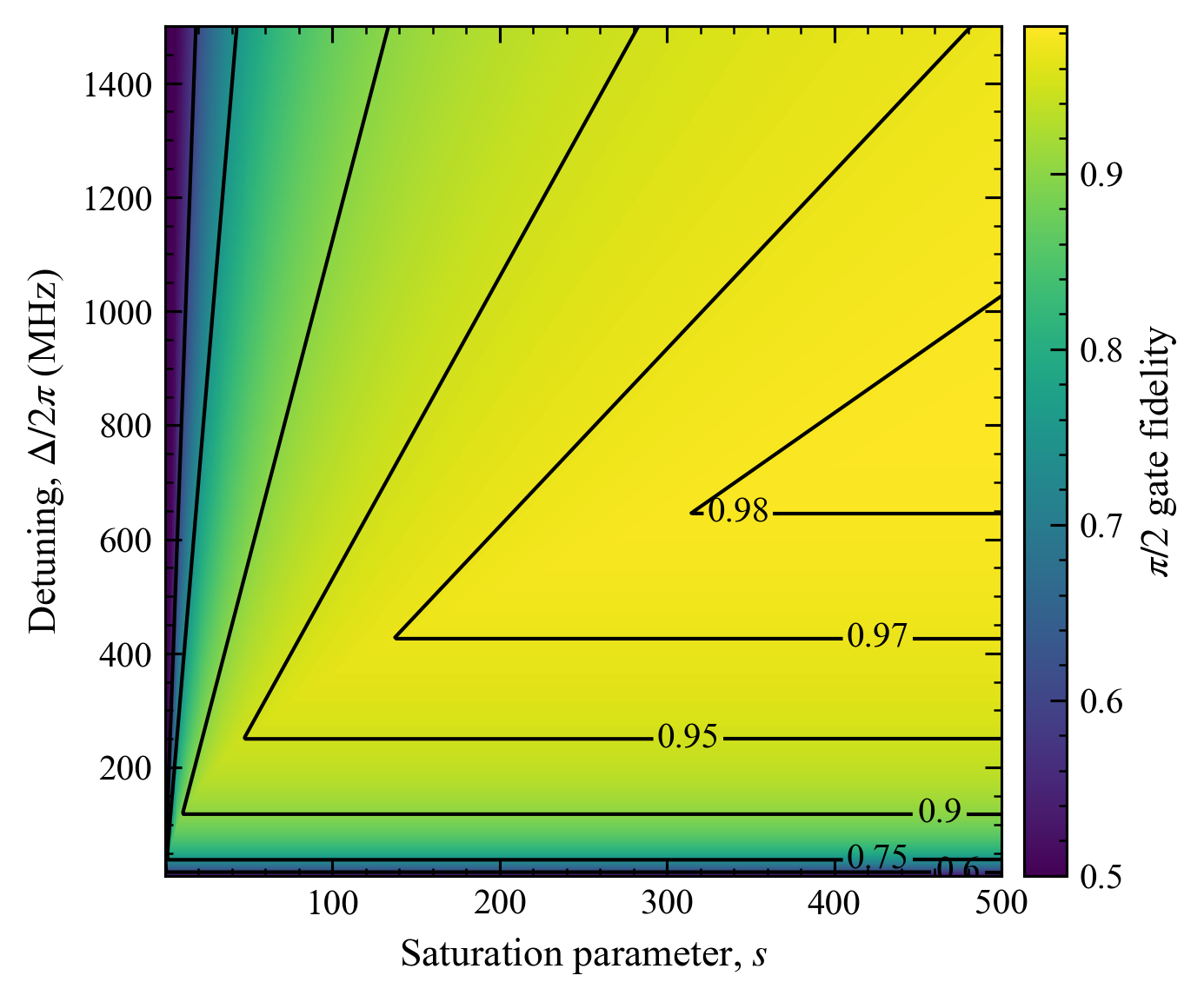}
\caption{\ensuremath{\pi}/2 gate fidelity as a function of saturation
parameter and single photon detuning. The contour lines display
parameter regions in which the fidelity is above the listed value. The
fidelity was calculated using the model described in the Raman Control
Section}
\label{fig:s3}
\end{figure}

\[H\  = \ \left( \frac{\Omega_{R}}{2} \right)\sigma_{x}\  + \ \left( \frac{\delta}{2} \right)\sigma_{z},\]

where \ensuremath{\Omega}\textsubscript{R} is the effective Raman Rabi frequency, \ensuremath{\delta} is the two-photon detuning, and \ensuremath{\sigma}\textsubscript{x} and \ensuremath{\sigma}\textsubscript{z} are Pauli spin operators. The \ensuremath{\sigma}\textsubscript{x} term drives coherent population transfer between the two spin states, while the \ensuremath{\sigma}\textsubscript{z} term accounts for an energy mismatch between the Raman drive and the ground-state splitting. Using the same Lindblad Master equation described above for the three-level model, we can define the collapse operators 
\[
C_{1}=\sqrt{\frac{\gamma_{1}}{2}}\,\sigma_{x},
\qquad
C_{2}=\sqrt{\frac{\gamma_{2}}{2}}\,\sigma_{z}, 
\]
in which $C_1$ represents population decay between the two states and $C_2$ represents pure dephasing. The characteristic relaxation and dephasing times were defined as
\[
T_{1}=\frac{1}{\gamma_{1}},
\qquad
T_{2,\mathrm{pure}}=\frac{1}{\gamma_{2}}.
\]

The initial density matrix was taken to be diagonal with all population initialized to the upper state. For each trial set of parameters, the master equation was numerically integrated over the experimental pulse durations. The lower spin-state population was obtained from
\[P_{0}(t)\  = \ Tr\lbrack|0\rangle\langle 0|\rho(t)\rbrack.\]

The principal fitted parameters were \ensuremath{\Omega}\textsubscript{R}, \ensuremath{\gamma}\textsubscript{1}, and \ensuremath{\gamma}\textsubscript{2}. To handle experimental noise and dark counts,, an additional scale factor and vertical offset were included:
\[S_{model}(t)\  = \ A P_{|0\rangle}(t)\  + \ B,\]

where A is the signal contrast and B is the baseline offset. The initial estimate for the optical scattering rate calculated above provided initial estimates the population decay and dephasing rates. The model was fit to the measured lower-state population using nonlinear least squares. This fit yielded \ensuremath{\Omega}/2\ensuremath{\pi} = 7.149 \ensuremath{\pm} 0.044 MHz, T\textsubscript{1} = 1.3(43) \ensuremath{\mu}s, and T\textsubscript{2,pure} = 0.26(33) \ensuremath{\mu}s. The measured oscillation frequency and scattering rates are therefore in reasonable agreement with the theoretical predictions. The large uncertainties for T\textsubscript{1} and $T_2$ arise because both \ensuremath{\gamma}\ensuremath{_1} and \ensuremath{\gamma}\ensuremath{_2} contribute to the damping of the oscillations. As a result, many different combinations of \ensuremath{\gamma}\ensuremath{_1} and \ensuremath{\gamma}\ensuremath{_2} can produce nearly the same decay envelope. The two parameters are therefore strongly correlated in the fit.

The Rabi measurement demonstrates coherent optical control and provides the \ensuremath{\pi}- and \ensuremath{\pi}/2-pulse durations used in the Ramsey, Hahn-echo, and CPMG measurements. Higher Raman Rabi frequencies and control fidelities could be achieved by increasing the optical power while simultaneously increasing the single-photon detuning to limit the corresponding rise in optical scattering. Such measurements would be better suited to emitters embedded in nanophotonic structures, where more efficient light coupling lowers the saturation power and therefore reduces the optical power required to similar fidelity pulses. 

\subsection{Ramsey Interferometry}
\begin{figure}[t]
\centering
\includegraphics[width=0.98\textwidth]{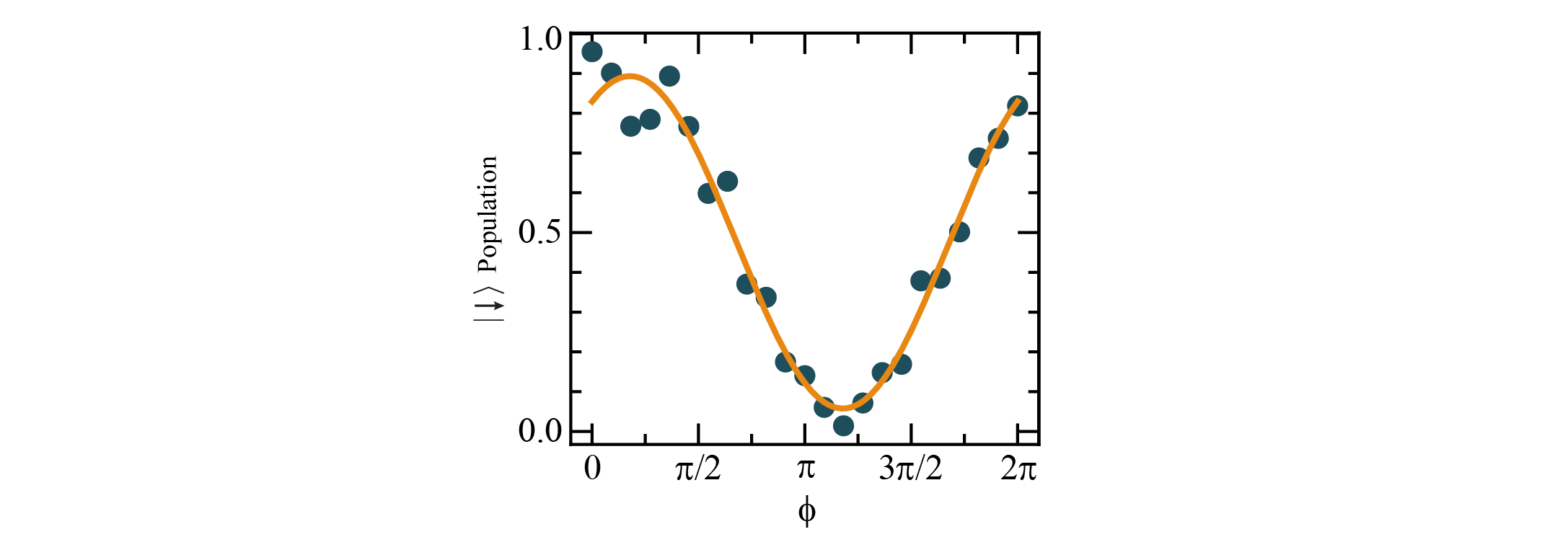}
\caption{Ramsey phase-sweep measurement at zero delay time. The phase of the second \ensuremath{\pi}/2 pulse was swept from 0 to 2\ensuremath{\pi}, and the measured spin-down population was fit to $P_{\downarrow}(\varphi) = a \cos(\varphi + \varphi_0) + b$. The fit yielded a = 0.418(18), b = 0.475(13), and $\varphi_0$ = −0.561(44) rad = −0.179(14)\ensuremath{\pi}, with $R^2$ = 0.965. The corresponding Ramsey-fringe visibility, calculated as V = a/b, was 0.880(45). }
\label{fig:s4}
\end{figure}

To perform Ramsey interferometry, we first extract the $\pi$ and $\pi/2$ pulse times from the fitted Rabi data. The first \ensuremath{\pi}/2 pulse prepared a coherent superposition of the two ground-state spin levels, after which the phase of the second \ensuremath{\pi}/2 pulse was swept from 0 to 2\ensuremath{\pi}. For ideal \ensuremath{\pi}/2 pulses, the probability of measuring the spin-down state following the application of the second $\pi/2$ pulse is
\[P_{\downarrow}(\varphi)\  = \ \frac{1}{2}\lbrack 1 + \mathrm{cos}(\varphi)\rbrack,\]

where $\varphi$ is the phase of the second \ensuremath{\pi}/2 pulse relative to the first. The measured population exhibited the expected cosine dependence as shown in Figure S4, with a slight phase shift which could be a result of a slight detuning off of the two photon resonance or an introduced phase shift from the AWG or MW amplifier.

To extract the inhomogeneous dephasing time, a Ramsey sequence consisting of two \ensuremath{\pi}/2 pulses separated by a free-evolution time, \ensuremath{\tau}, was used. The first $\pi/2$ pulse had a phase of $0^\circ$ but the second $\pi/2$ pulse had a phase that evolved based on how long the free-evolution time was. In particular, it was advanced according to \[\varphi\  = \ \omega_{S}\tau,\]
where $\omega_{S}$ is an artificially imposed phase-ramp frequency. This converted the phase that had accumulated during the free evolution time into a measurable population difference. For these measurements, $\omega_{S}/2\pi=15$ MHz. In the absence of two-photon detuning and differential AC Stark shifts, the measured Ramsey frequency would equal the applied phase-ramp frequency: $\omega_{\mathrm{Ramsey}} = \omega_{S}$. More generally, the
observed Ramsey frequency is
\[\omega_{\mathrm{Ramsey}}\  = \ \omega_{S} + \delta + \Delta_{AC},\]

where \ensuremath{\delta} is the two-photon detuning and $\Delta_{\mathrm{AC}}$ is the differential AC Stark shift produced during the Raman pulses. The differential AC Stark shift arises because the two ground states do not experience identical optical energy shifts. Although the optical fields are absent during the free-evolution interval, the spin precesses relative to the rotating frame established by the driven Raman pulses. The Ramsey data associated with this measurement was fit using a sinusoid with an exponential decay envelope:
\[P_{\downarrow}(\tau)\  = \ a\ \exp\lbrack - \tau/T_{2}^{*}\ \rbrack\ \mathrm{sin}(\omega_{\mathrm{Ramsey}}\tau\  + \ \alpha)\  + \ b,\]

where a is the oscillation amplitude, b is the population offset, \ensuremath{\alpha} is the phase offset, and $T_2^*$ is the inhomogeneous dephasing time. The fit yielded $T_{2}^{*} =  371.7 \pm 53.6\ \mathrm{ns}$ and $\omega_{\mathrm{Ramsey}}/2\pi = 12.59 \pm 0.062$ MHz. The measured Ramsey frequency was lower than the imposed phase-ramp frequency of 15 MHz. Under the sign convention used above, the combined contribution from the two-photon detuning and differential AC Stark shift was therefore $(\delta + \Delta_{AC})/2\pi = - 2.41 \pm 0.062\ $ MHz. A single Ramsey measurement cannot independently distinguish the contribution of the two-photon detuning from that of the differential AC Stark shift. Separating these effects would require measuring the Ramsey frequency as a function of the applied two-photon detuning.

\subsection{Hahn-echo and CPMG sequences}
To extract the visibility for the Hahn-echo and CPMG sequences, two phases for the final p/2 pulse were used with \ensuremath{\phi} = 0 and \ensuremath{\phi}~= \ensuremath{\pi} as these should correspond to maxima and minima. Then, the visibility can be calculated as
\[V(\tau)\  = \ \lbrack S_{\max}(\tau) - S_{\min}(\tau)\rbrack/\lbrack S_{\max}(\tau) + S_{\min}(\tau)\rbrack,\]

where S\textsubscript{max} and S\textsubscript{min} are the background-subtracted ratio between readout and initialization signals measured at the two final-pulse phases. Pulse imperfections can reduce the measured visibility. These may arise from optical scattering, small errors in the $\pi$- or $\pi/2$-pulse durations, pulse-area fluctuations caused by optical-power instability, or detuning from the two-photon resonance. This will not change the stretch exponent nor the decay time extracted, it simply changes the baselines measured. Nevertheless, the visibility generally decreases for longer pulse sequences because the larger number of pulses provides more opportunities for pulse errors to accumulate. The slightly higher baseline visibility observed for CPMG-4 compared with CPMG-2 results from improvements made to the experimental setup after the CPMG-2 measurement. In particular, we reduced the detuning from the two-photon resonance and more carefully calibrated the $\pi$- and $\pi/2$-pulse durations. These improvements increased the signal-to-noise ratio and made small, nonzero visibilities easier to resolve.

\begin{figure}[t]
\centering
\includegraphics[width=0.7\textwidth]{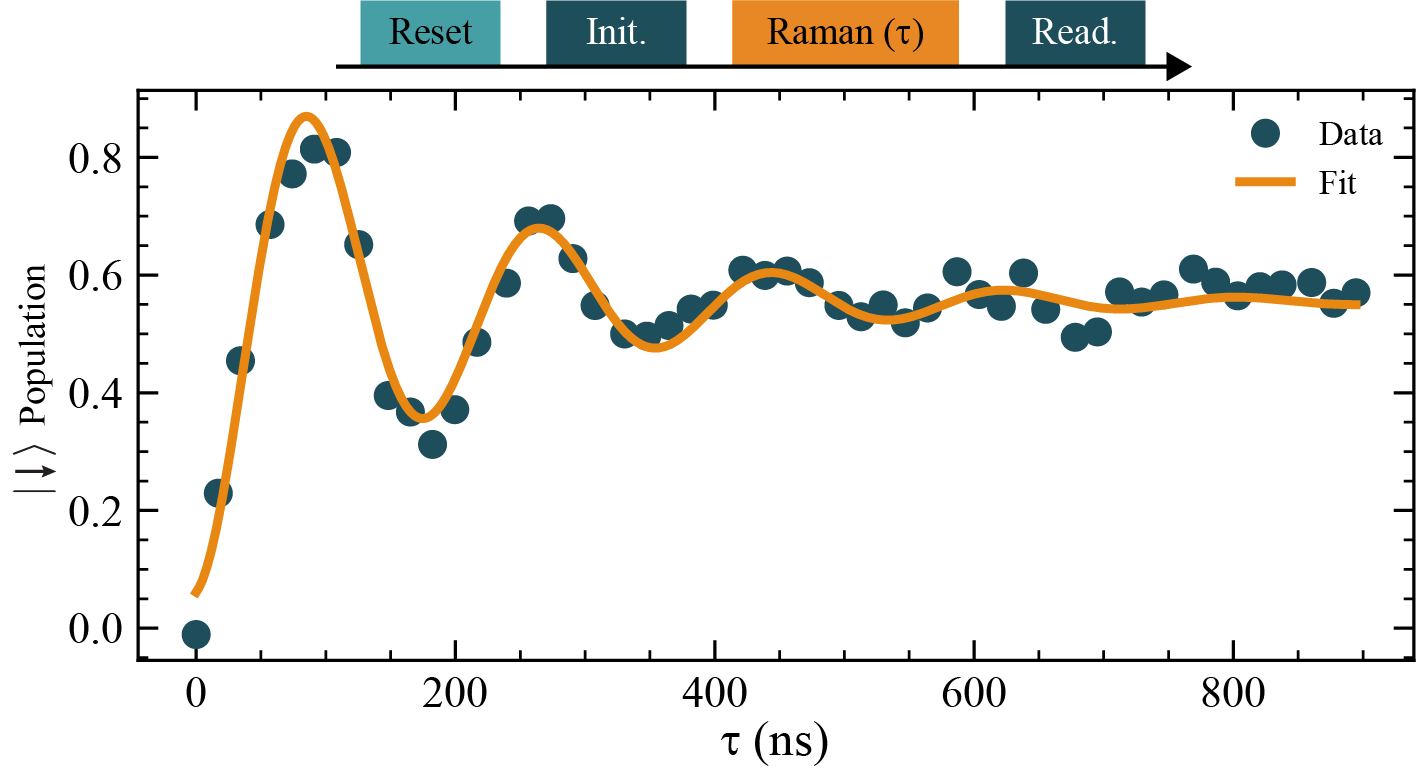}
\caption{Rabi data taken with 14 $\mu$W of power at 40 mT. The data is fitted using the two level model described above with a T\ensuremath{_1} = 0.26 (17) \ensuremath{\mu}s and T\ensuremath{_2},\textsubscript{pure} = 0.42 \ensuremath{\pm} 0.85 \ensuremath{\mu}s extracted. The
shorter than expected T1 time is potentially consistent with additional
cross-excitation from a higher order sideband. This could be mitigated
by further suppressing such higher order sidebands or placing the
carrier further away, so that the generated sidebands do not overlap
with the transitions originating from the higher excited $\ket{B}$ state transitions.}
\label{fig:s5}
\end{figure}

\subsection{Low Magnetic Field Rabi}
This measurement is a Rabi measurement taken at 40 mT using the same single photon detuning and power before the objective. The measured Rabi rate is 5.621 $\pm$ 0.084 MHz, close to what was measured at 150 mT, demonstrating that spin control can be achieved on similar times scales at low magnetic field and the excited state spin quantization axis angle is not a function of magnetic field strength. Thus \ensuremath{\eta}~is similar at low magnetic fields and spin control can be just as efficient. To measure this however, the carrier frequency was placed at a lower
frequency compared to both the A2 and A1 transitions rather than in
between both transitions. In particular, it was placed such that the
frequency spacing between the carrier and A2 was equal to the splitting
between A1 and A2 and the single photon detuning. Then, for the Raman
drive, a two-tone sin wave was used with the splitting frequency and
twice the splitting frequency used as the two sin frequencies, which
were then combined and sent to the EOM. This is likely the reason for the slightly slower Rabi frequency despite the same measured power before the objective lens. Namely, the modulation efficiency of the EOM for the two-tone Raman drive is less efficient compared to the single sin-wave generated before. As such, less power is going into the Raman drive compared to before. Importantly, however, this modulation efficiency is not a fundamental limit and can be improved. The reset and initialization pulses were also performed using sin waves instead of a
serrodyne scheme, opening up the possibility for using simpler RF
sources for control instead of a costly AWG. 

\subsection{Autocorrelation g$^{(2)}(\tau)$ measurement}

To confirm that the emitter we were working with was in fact a single
NiV, we measured the second-order photon autocorrelation using a
Hanbury-Brown and Twiss setup. We use a background correction technique
to account for the detector dark counts
\cite{ref19}
and the characteristic $g^{(2)}$ is shown in Figure S6. The
fitted line is a coherent model that includes a damped sin to account
for resonant Rabi oscillations between the ground and excited state as
described previously in
\cite{ref42}.
The measured minimum value is 0.003 consistent with a single emitter.

    \begin{figure}[h]
    \centering
    \includegraphics[width=0.4\columnwidth]{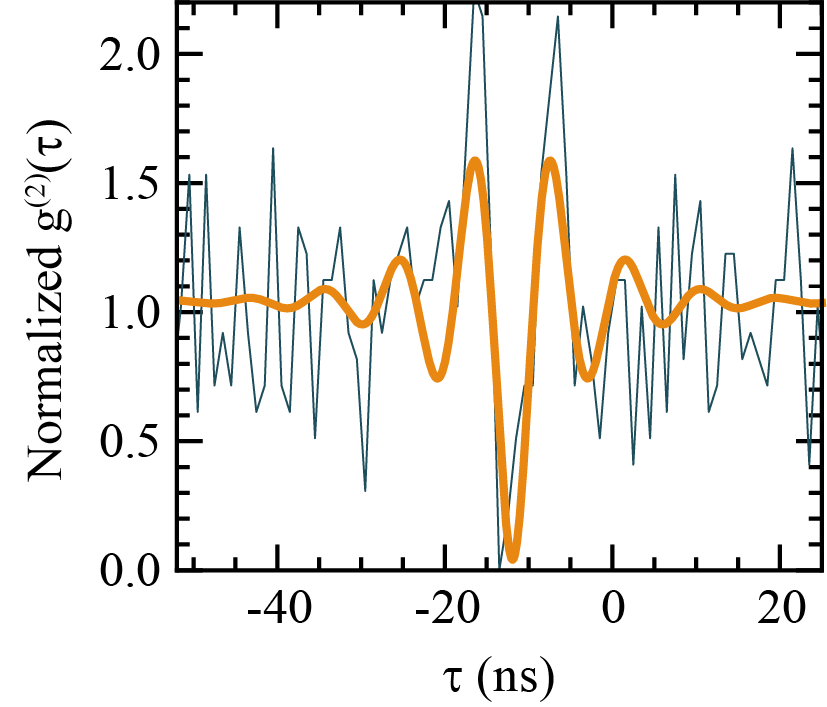}
    \caption{Autocorrelation measurement. A coherent $g^{2}(t)$
function was used to fit the data as described in the main section. The
minimum point is shifted because of timing differences in the two
detectors and counting channels. We find $g^{(2)}(0)$ = 0.003
consistent with a single emitter.}
    \label{fig:s6}
    \end{figure}


\begin{thebibliography}{0}%
\makeatletter
\providecommand \@ifxundefined [1]{%
 \@ifx{#1\undefined}
}%
\providecommand \@ifnum [1]{%
 \ifnum #1\expandafter \@firstoftwo
 \else \expandafter \@secondoftwo
 \fi
}%
\providecommand \@ifx [1]{%
 \ifx #1\expandafter \@firstoftwo
 \else \expandafter \@secondoftwo
 \fi
}%
\providecommand \natexlab [1]{#1}%
\providecommand \enquote  [1]{``#1''}%
\providecommand \bibnamefont  [1]{#1}%
\providecommand \bibfnamefont [1]{#1}%
\providecommand \citenamefont [1]{#1}%
\providecommand \href@noop [0]{\@secondoftwo}%
\providecommand \href [0]{\begingroup \@sanitize@url \@href}%
\providecommand \@href[1]{\@@startlink{#1}\@@href}%
\providecommand \@@href[1]{\endgroup#1\@@endlink}%
\providecommand \@sanitize@url [0]{\catcode `\\12\catcode `\$12\catcode `\&12\catcode `\#12\catcode `\^12\catcode `\_12\catcode `\%12\relax}%
\providecommand \@@startlink[1]{}%
\providecommand \@@endlink[0]{}%
\providecommand \url  [0]{\begingroup\@sanitize@url \@url }%
\providecommand \@url [1]{\endgroup\@href {#1}{\urlprefix }}%
\providecommand \urlprefix  [0]{URL }%
\providecommand \Eprint [0]{\href }%
\providecommand \doibase [0]{https://doi.org/}%
\providecommand \selectlanguage [0]{\@gobble}%
\providecommand \bibinfo  [0]{\@secondoftwo}%
\providecommand \bibfield  [0]{\@secondoftwo}%
\providecommand \translation [1]{[#1]}%
\providecommand \BibitemOpen [0]{}%
\providecommand \bibitemStop [0]{}%
\providecommand \bibitemNoStop [0]{.\EOS\space}%
\providecommand \EOS [0]{\spacefactor3000\relax}%
\providecommand \BibitemShut  [1]{\csname bibitem#1\endcsname}%
\let\auto@bib@innerbib\@empty
\end{thebibliography}%


\begin{thebibliography}{50}

\bibitem{ref1} H. J. Kimble, The quantum internet. \emph{Nature}. \textbf{453}, 1023--1030
(2008).

\bibitem{ref2} S. Wehner, D. Elkouss, R. Hanson, Quantum internet: A vision for the road ahead.
\emph{Science}. \textbf{362} (2018), doi:10.1126/science.aam9288.

\bibitem{ref3} A. K. Ekert, Quantum cryptography based on Bell's theorem. \emph{Physical review letters}. \textbf{67}, 661 (1991).

\bibitem{ref4} N. Gisin, G. Ribordy, W. Tittel, H. Zbinden, Quantum cryptography. \emph{Rev. Mod. Phys.} \textbf{74}, 145--195 (2002).

\bibitem{ref5} C. Monroe, R. Raussendorf, A. Ruthven, K. R. Brown, P. Maunz, L. M. Duan, J. Kim, Large-scale modular quantum-computer architecture with atomic memory and photonic interconnects. \emph{Phys. Rev. A}. \textbf{89}, 022317 (2014).

\bibitem{ref6} N. H. Nickerson, J. F. Fitzsimons, S. C. Benjamin, Freely scalable quantum technologies using cells of 5-to-50 qubits with very lossy and noisy photonic links. \emph{Physical Review X}. \textbf{4}, 041041 (2014).

\bibitem{ref7} D. Gottesman, T. Jennewein, S. Croke, Longer-baseline telescopes using quantum repeaters. \emph{Phys. Rev. Lett.} \textbf{109}, 070503 (2012).

\bibitem{ref8} P. Kómár, E. M. Kessler, M. Bishof, L. Jiang, A. S. Sørensen, J. Ye, M. D. Lukin, A quantum network of clocks. \emph{Nat. Phys.} \textbf{10}, 582--587 (2014).

\bibitem{ref9} M. Ruf, N. H. Wan, H. Choi, D. Englund, R. Hanson, Quantum networks based on color centers in diamond. \emph{J. Appl. Phys.} \textbf{130}, 070901 (2021).

\bibitem{ref10} E. Togan, Y. Chu, A. S. Trifonov, L. Jiang, J. Maze, L. Childress, M. V. G. Dutt, A. S. Sørensen, P. R. Hemmer, A. S. Zibrov, M. D. Lukin, Quantum entanglement between an optical photon and a solid-state spin qubit. \emph{Nature}. \textbf{466}, 730--734 (2010).

\bibitem{ref11} B. Hensen, H. Bernien, A. E. Dréau, A. Reiserer, N. Kalb, M. S. Blok, J. Ruitenberg, R. F. L. Vermeulen, R. N. Schouten, C. Abellán, W. Amaya, V. Pruneri, M. W. Mitchell, M. Markham, D. J. Twitchen, D. Elkouss, S. Wehner, T. H. Taminiau, R. Hanson, Loophole-free Bell inequality violation using electron spins separated by 1.3 kilometres. \emph{Nature}. \textbf{526}, 682--686 (2015).

\bibitem{ref12} M. Pompili, S. L. N. Hermans, S. Baier, H. K. C. Beukers, P. C. Humphreys, R. N. Schouten, R. F. L. Vermeulen, M. J. Tiggelman, L. Dos Santos Martins, B. Dirkse, S. Wehner, R. Hanson, Realization of a multinode quantum network of remote solid-state qubits. \emph{Science}. \textbf{372}, 259--264 (2021).

\bibitem{ref13} A. J. Stolk, K. L. van der Enden, M.-C. Slater, I. Te Raa-Derckx, P. Botma, J. van Rantwijk, J. J. B. Biemond, R. A. J. Hagen, R. W. Herfst, W. D. Koek, A. J. H. Meskers, R. Vollmer, E. J. van Zwet, M. Markham, A. M. Edmonds, J. F. Geus, F. Elsen, B. Jungbluth, C. Haefner, C. Tresp, J. Stuhler, S. Ritter, R. Hanson, Metropolitan-scale heralded entanglement of solid-state qubits. \emph{Sci. Adv.} \textbf{10}, eadp6442 (2024).

\bibitem{ref14} L. Childress, R. Hanson, Diamond NV centers for quantum computing and quantum networks. \emph{MRS Bull.} \textbf{38}, 134--138 (2013).

\bibitem{ref15} D. Riedel, I. Söllner, B. J. Shields, S. Starosielec, P. Appel, E. Neu, P. Maletinsky, R. J. Warburton, Deterministic Enhancement of Coherent Photon Generation from a Nitrogen-Vacancy Center in Ultrapure Diamond. \emph{Phys. Rev. X}. \textbf{7}, 031040 (2017).

\bibitem{ref16} P. Tamarat, T. Gaebel, J. R. Rabeau, M. Khan, A. D. Greentree, H. Wilson, L. C. L. Hollenberg, S. Prawer, P. Hemmer, F. Jelezko, J. Wrachtrup, Stark shift control of single optical centers in diamond. \emph{Phys. Rev. Lett.} \textbf{97}, 083002 (2006).

\bibitem{ref17} A. Batalov, C. Zierl, T. Gaebel, P. Neumann, I. Y. Chan, G. Balasubramanian, P. R. Hemmer, F. Jelezko, J. Wrachtrup, Temporal coherence of photons emitted by single nitrogen-vacancy defect centers in diamond using optical Rabi-oscillations. \emph{Phys. Rev. Lett.} \textbf{100}, 077401 (2008).

\bibitem{ref18} H. Bernien, B. Hensen, W. Pfaff, G. Koolstra, M. S. Blok, L. Robledo, T. H. Taminiau, M. Markham, D. J. Twitchen, L. Childress, R. Hanson, Heralded entanglement between solid-state qubits separated by three metres. \emph{Nature}. \textbf{497}, 86--90 (2013).

\bibitem{ref19} E. Neu, M. Agio, C. Becher, Photophysics of single silicon vacancy centers in diamond: implications for single photon emission. \emph{Opt. Express}. \textbf{20}, 19956--19971 (2012).

\bibitem{ref20} C. Hepp, Electronic structure of the silicon vacancy color center in diamond. \emph{Universität des Saarlandes} (2014), doi:10.22028/d291-23020.

\bibitem{ref21} L. J. Rogers, K. D. Jahnke, M. H. Metsch, A. Sipahigil, J. M. Binder, T. Teraji, H. Sumiya, J. Isoya, M. D. Lukin, P. Hemmer, F. Jelezko, All-optical initialization, readout, and coherent preparation of single silicon-vacancy spins in diamond. \emph{Phys. Rev. Lett.} \textbf{113}, 263602 (2014).

\bibitem{ref22} A. Sipahigil, K. D. Jahnke, L. J. Rogers, T. Teraji, J. Isoya, A. S.
Zibrov, F. Jelezko, M. D. Lukin, Indistinguishable photons from separated silicon-vacancy centers in diamond. \emph{Phys. Rev. Lett.} \textbf{113}, 113602 (2014).

\bibitem{ref23} C. Hepp, T. Müller, V. Waselowski, J. N. Becker, B. Pingault, H. Sternschulte, D. Steinmüller-Nethl, A. Gali, J. R. Maze, M. Atatüre, C. Becher, Electronic structure of the silicon vacancy color center in diamond. \emph{Phys. Rev. Lett.} \textbf{112}, 036405 (2014).

\bibitem{ref24} A. Faraon, C. Santori, Z. Huang, V. M. Acosta, R. G. Beausoleil, Coupling of nitrogen-vacancy centers to photonic crystal cavities in monocrystalline diamond. \emph{Phys. Rev. Lett.} \textbf{109}, 033604 (2012).

\bibitem{ref25} P. Siyushev, H. Pinto, M. Vörös, A. Gali, F. Jelezko, J. Wrachtrup, Optically controlled switching of the charge state of a single nitrogen-vacancy center in diamond at cryogenic temperatures. \emph{Phys. Rev. Lett.} \textbf{110}, 167402 (2013).

\bibitem{ref26} B. Pingault, J. N. Becker, C. H. H. Schulte, C. Arend, C. Hepp, T. Godde, A. I. Tartakovskii, M. Markham, C. Becher, M. Atatüre, All-optical formation of coherent dark states of silicon-vacancy spins in diamond. \emph{Phys. Rev. Lett.} \textbf{113}, 263601 (2014).

\bibitem{ref27} K. D. Jahnke, A. Sipahigil, J. M. Binder, M. W. Doherty, M. Metsch, L. J. Rogers, N. B. Manson, M. D. Lukin, F. Jelezko, Electron--phonon processes of the silicon-vacancy centre in diamond. \emph{New J. Phys.} \textbf{17}, 043011 (2015).

\bibitem{ref28} J. N. Becker, B. Pingault, D. Groß, M. Gündoğan, N. Kukharchyk, M. Markham, A. Edmonds, M. Atatüre, P. Bushev, C. Becher, All-Optical Control of the Silicon-Vacancy Spin in Diamond at Millikelvin Temperatures. \emph{Phys. Rev. Lett.} \textbf{120}, 053603 (2018).

\bibitem{ref29} D. D. Sukachev, A. Sipahigil, C. T. Nguyen, M. K. Bhaskar, R. E. Evans, F. Jelezko, M. D. Lukin, Silicon-Vacancy Spin Qubit in Diamond: A Quantum Memory Exceeding 10~ms with Single-Shot State Readout. \emph{Phys. Rev. Lett.} \textbf{119}, 223602 (2017).

\bibitem{ref30} S. Meesala, Y.-I. Sohn, B. Pingault, L. Shao, H. A. Atikian, J. Holzgrafe, M. Gündoğan, C. Stavrakas, A. Sipahigil, C. Chia, R. Evans, M. J. Burek, M. Zhang, L. Wu, J. L. Pacheco, J. Abraham, E. Bielejec, M. D. Lukin, M. Atatüre, M. Lončar, Strain engineering of the silicon-vacancy center in diamond. \emph{Phys. Rev. B}. \textbf{97}, 205444 (2018).

\bibitem{ref31} E. Bersin, M. Sutula, Y. Q. Huan, A. Suleymanzade, D. R. Assumpcao, Y.-C. Wei, P.-J. Stas, C. M. Knaut, E. N. Knall, C. Langrock, N. Sinclair, R. Murphy, R. Riedinger, M. Yeh, C. J. Xin, S. Bandyopadhyay, D. D. Sukachev, B. Machielse, D. S. Levonian, M. K. Bhaskar, S. Hamilton, H. Park, M. Lončar, M. M. Fejer, P. B. Dixon, D. R. Englund, M. D. Lukin, Telecom Networking with a Diamond Quantum Memory. \emph{PRX Quantum}. \textbf{5}, 010303 (2024).

\bibitem{ref32} B. C. Rose, D. Huang, Z.-H. Zhang, P. Stevenson, A. M. Tyryshkin, S. Sangtawesin, S. Srinivasan, L. Loudin, M. L. Markham, A. M. Edmonds, D. J. Twitchen, S. A. Lyon, N. P. de Leon, Observation of an environmentally insensitive solid-state spin defect in diamond. \emph{Science}. \textbf{361}, 60--63 (2018).

\bibitem{ref33} Z.-H. Zhang, J. A. Zuber, L. V. H. Rodgers, X. Gui, P. Stevenson, M. Li, M. Batzer, M. L. Grimau Puigibert, B. J. Shields, A. M. Edmonds, N. Palmer, M. L. Markham, R. J. Cava, P. Maletinsky, N. P. de Leon, Neutral silicon vacancy centers in undoped diamond via surface control. \emph{Phys. Rev. Lett.} \textbf{130}, 166902 (2023).

\bibitem{ref34} B. L. Green, S. Mottishaw, B. G. Breeze, A. M. Edmonds, U. F. S. D'Haenens-Johansson, M. W. Doherty, S. D. Williams, D. J. Twitchen, M. E. Newton, Neutral Silicon-Vacancy Center in Diamond: Spin Polarization and Lifetimes. \emph{Phys. Rev. Lett.} \textbf{119}, 096402 (2017).

\bibitem{ref35} T. Iwasaki, F. Ishibashi, Y. Miyamoto, Y. Doi, S. Kobayashi, T. Miyazaki, K. Tahara, K. D. Jahnke, L. J. Rogers, B. Naydenov, F. Jelezko, S. Yamasaki, S. Nagamachi, T. Inubushi, N. Mizuochi, M. Hatano, Germanium-Vacancy Single Color Centers in Diamond. \emph{Sci. Rep.} \textbf{5}, 12882 (2015).

\bibitem{ref36} T. Iwasaki, Y. Miyamoto, T. Taniguchi, P. Siyushev, M. H. Metsch, F. Jelezko, M. Hatano, Tin-Vacancy Quantum Emitters in Diamond. \emph{Phys. Rev. Lett.} \textbf{119}, 253601 (2017).

\bibitem{ref37} M. E. Trusheim, N. H. Wan, K. C. Chen, C. J. Ciccarino, J. Flick, R.
Sundararaman, G. Malladi, E. Bersin, M. Walsh, B. Lienhard, H. Bakhru, P. Narang, D. Englund, Lead-related quantum emitters in diamond. \emph{Phys. Rev. B}. \textbf{99}, 075430 (2019).

\bibitem{ref38} G. Thiering, A. Gali, \emph{Ab~Initio} Magneto-Optical Spectrum of Group-IV Vacancy Color Centers in Diamond. \emph{Phys. Rev. X}. \textbf{8}, 021063 (2018).

\bibitem{ref39} R. Debroux, C. P. Michaels, C. M. Purser, N. Wan, M. E. Trusheim, J. Arjona Martínez, R. A. Parker, A. M. Stramma, K. C. Chen, L. de Santis, E. M. Alexeev, A. C. Ferrari, D. Englund, D. A. Gangloff, M. Atatüre, Quantum Control of the Tin-Vacancy Spin Qubit in Diamond. \emph{Phys. Rev. X}. \textbf{11}, 041041 (2021).

\bibitem{ref40} E. I. Rosenthal, C. P. Anderson, H. C. Kleidermacher, A. J. Stein, H. Lee, J. Grzesik, G. Scuri, A. E. Rugar, D. Riedel, S. Aghaeimeibodi, G. H. Ahn,  K. Van Gasse, J. Vučković, Microwave Spin Control of a Tin-Vacancy Qubit in Diamond. \emph{Phys. Rev. X}. \textbf{13}, 031022 (2023).

\bibitem{ref41} I. Karapatzakis, J. Resch, M. Schrodin, P. Fuchs, M. Kieschnick, J. Heupel, L. Kussi, C. Sürgers, C. Popov, J. Meijer, C. Becher, W. Wernsdorfer, D. Hunger, Microwave Control of the Tin-Vacancy Spin Qubit in Diamond with a Superconducting Waveguide. \emph{Phys. Rev. X}. \textbf{14}, 031036
(2024).

\bibitem{ref42} I. M. Morris, T. Lühmann, K. Klink, L. Crooks, D. Hardeman, D. J. Twitchen, S. Pezzagna, J. Meijer, S. S. Nicley, J. N. Becker, Lifetime-Limited and Tunable Emission from Single Charge-Stabilized Nickel Vacancy Centers in Diamond. \emph{Phys. Rev. Lett.} \textbf{135}, 043602 (2025).

\bibitem{ref43} G. Thiering, A. Gali, Magneto-optical spectra of the split nickel-vacancy defect in diamond. \emph{Phys. Rev. Research}. \textbf{3}, 043052
(2021).

\bibitem{ref44} E. L. Hahn, Spin Echoes. \emph{Phys. Rev.} \textbf{80}, 580--594 (1950).

\bibitem{ref45} U. Cywinski, R. M. Lutchyn, C. P. Nave, S. Das Sarma, How to enhance
dephasing time in superconducting qubits. \emph{Physical Review B---Condensed Matter and Materials Physics}. \textbf{77}, 174509
(2008).

\bibitem{ref46} J. R. Maze, P. L. Stanwix, J. S. Hodges, S. Hong, J. M. Taylor, P. Cappellaro, L. Jiang, M. V. G. Dutt, E. Togan, A. S. Zibrov, A. Yacoby, R. L.
Walsworth, M. D. Lukin, Nanoscale magnetic sensing with an individual electronic spin in diamond. \emph{Nature}. \textbf{455}, 644--647 (2008).

\bibitem{ref47} G. de Lange, Z. H. Wang, D. Ristè, V. V. Dobrovitski, R. Hanson, Universal dynamical decoupling of a single solid-state spin from a spin bath. \emph{Science}. \textbf{330}, 60--63 (2010).

\bibitem{ref48} H. Y. Carr, E. M. Purcell, Effects of Diffusion on Free Precession in Nuclear Magnetic Resonance Experiments. \emph{Phys. Rev.} \textbf{94}, 630--638 (1954).

\bibitem{ref49} S. Meiboom, D. Gill, Modified Spin-Echo Method for Measuring Nuclear Relaxation Times. \emph{Rev. Sci. Instrum.} \textbf{29}, 688 (1958).

\bibitem{ref50} C. J. Foot, \emph{Atomic physics} (Oxford university press, 2005), vol. 7.

\end{thebibliography}
\end{document}